\pdfoutput=1
\documentclass[cernpreprint, atlasdraft=false, texlive=2023, UKenglish, texmf, orcidlogo]{atlasdoc}
 
\usepackage{atlaspackage}
\usepackage{atlasbiblatex}
 
\usepackage{atlasphysics}
 
\usepackage{multirow}
 
\graphicspath{{logos/}{figures/}}
 
\usepackage{ANA-STDM-2018-59-PAPER-defs}

\AtlasTitle{Measurement of electroweak $Z(\nu\bar{\nu})\gamma jj$ production and limits on anomalous quartic gauge couplings in $pp$ collisions at $\sqrt{s}=13$~TeV with the ATLAS detector}

\AtlasAbstract{
The electroweak production of $Z(\nu\bar{\nu})\gamma$ in association with two jets is studied in a regime with a photon of high transverse momentum above 150~GeV using proton--proton collisions at a centre-of-mass energy of 13~TeV at the Large Hadron Collider. The analysis uses a data sample with an integrated luminosity of 139~fb$^{-1}$ collected by the ATLAS detector during the 2015--2018 LHC data-taking period.
This process is an important probe of the electroweak symmetry breaking mechanism in the Standard Model and is sensitive to quartic gauge boson couplings via vector-boson scattering. The fiducial $Z(\nu\bar{\nu})\gamma jj$ cross section for electroweak production is measured to be 0.77$^{+0.34}_{-0.30}$~fb and is consistent with the Standard Model prediction.
Evidence of electroweak $Z(\nu\bar{\nu})\gamma jj$ production is found with an observed significance of 3.2$\sigma$ for the background-only hypothesis, compared with an expected significance of 3.7$\sigma$.
The combination of this result with the previously published ATLAS observation of electroweak $Z(\nu\bar{\nu})\gamma jj$ production yields an observed (expected) signal significance of 6.3$\sigma$~(6.6$\sigma$). Limits on anomalous quartic gauge boson couplings are obtained in the framework of effective field theory with dimension-8 operators.
}

\AtlasRefCode{STDM-2018-59}
 
\PreprintIdNumber{CERN-EP-2022-138}

\AtlasJournal{JHEP}
\AtlasJournalRef{\JHEP 06 (2023) 082}
\AtlasDOI{10.1007/JHEP06(2023)082}

\hypersetup{pdftitle={ATLAS document},pdfauthor={The ATLAS Collaboration}}
 
\newcommand{\Zgjj}{\ensuremath{Z \gamma jj}\xspace}

\newcommand{\mjj}{\ensuremath{m_{jj}}}
\newcommand{\metpt}{\ensuremath{\vec{p}_{\mathrm{T}}^{\,\text{miss}}}}
\newcommand{\etogam}{\ensuremath{e\rightarrow\gamma}\xspace}
\newcommand{\jtogam}{\ensuremath{j\rightarrow\gamma}\xspace}
\newcommand{\ETmiss}{\ensuremath{E_\mathrm{T}^{\mathrm{miss}}}\xspace}
\newcommand{\gj}{\ensuremath{\gamma\text{+jets}}\xspace}

\begin{document}
 
\maketitle
 
\tableofcontents
 
\section{Introduction}
\label{sec:intro}

 
The scattering of two vector bosons (VBS), $V V \to V V$ with $V$ = $W$/$Z$/$\gamma$, is an important process to probe the nature of electroweak (EWK) symmetry breaking in the Standard Model (SM). The presence of the Higgs field in the SM prevents the divergence of VBS amplitudes at high energies and violation of unitarity at the~\TeV~scale. The non-Abelian structure of gauge interactions in the electroweak sector of the Standard Model results in a rich variety of VBS processes, with unique opportunities to probe physics beyond the SM (BSM). VBS processes are sensitive to SM quartic gauge couplings (QGCs) and also to possible anomalous QGCs (aQGCs)~\cite{Eboli:2003nq,PhysRevD.93.093013,Baak:2013fwa}.

Neutral QGCs are absent in the SM at tree level, but they can be induced by BSM physics, allowing new processes at high energy scales. The vector-boson scattering $Z\gamma$ process and a similar process exploiting the $ZZ$ final state are the only processes sensitive to the neutral quartic gauge couplings, with the former having the larger expected cross section. Such processes can be studied through measurements of the electroweak production of two vector bosons and two jets ($VVjj$).
 
\begin{figure}[h!]
\begin{center}
\includegraphics[width=0.33\columnwidth]{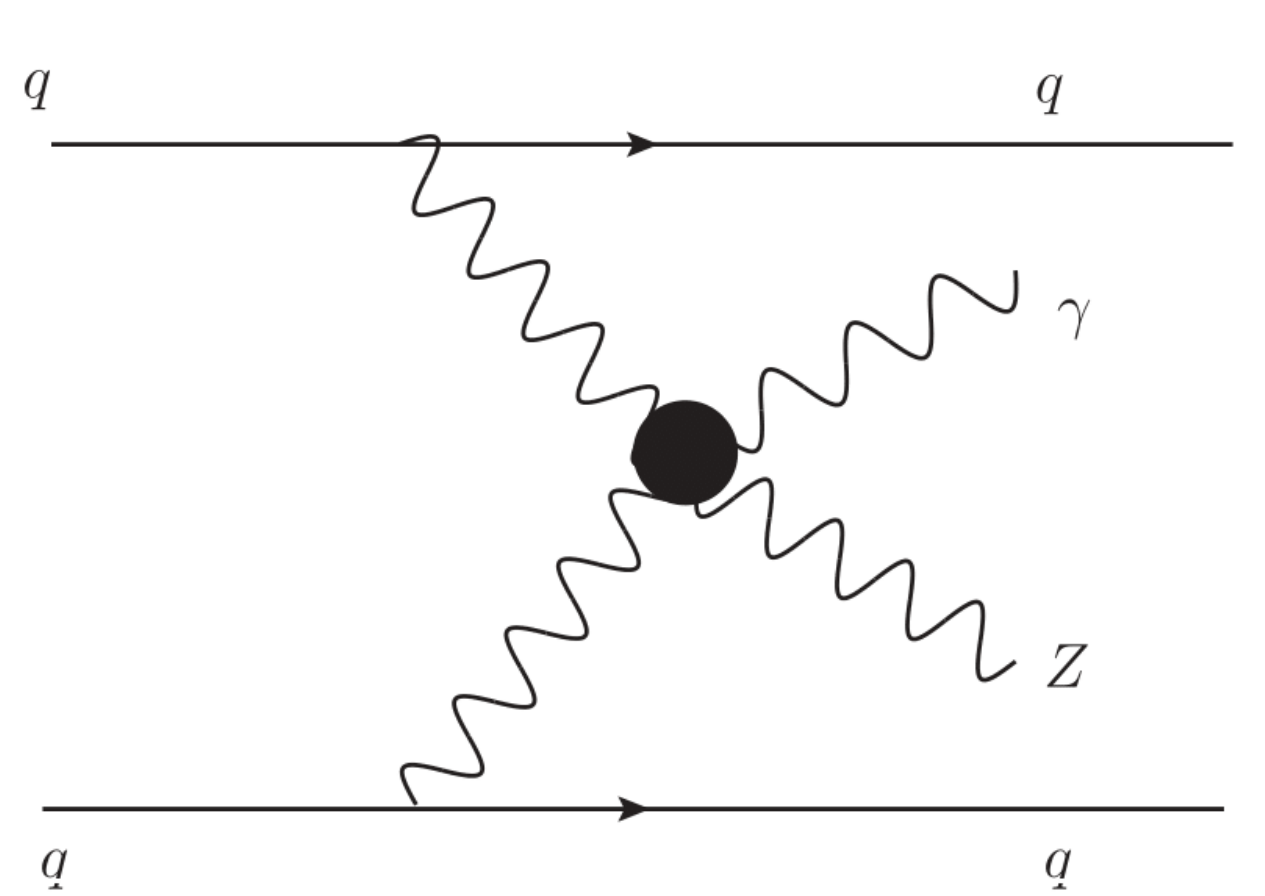} \hspace{1cm}
\includegraphics[width=0.33\columnwidth]{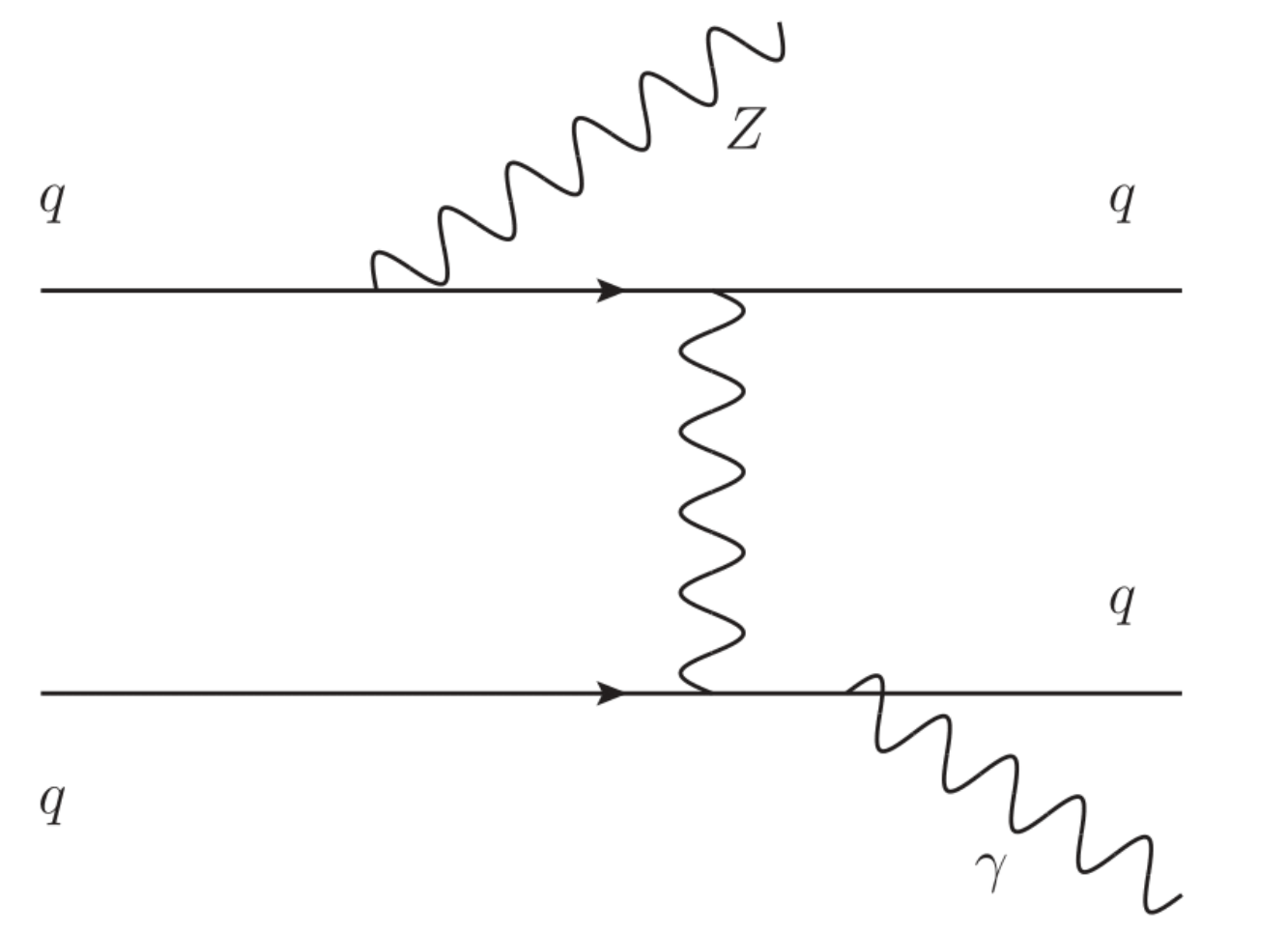}
\includegraphics[width=0.33\columnwidth]{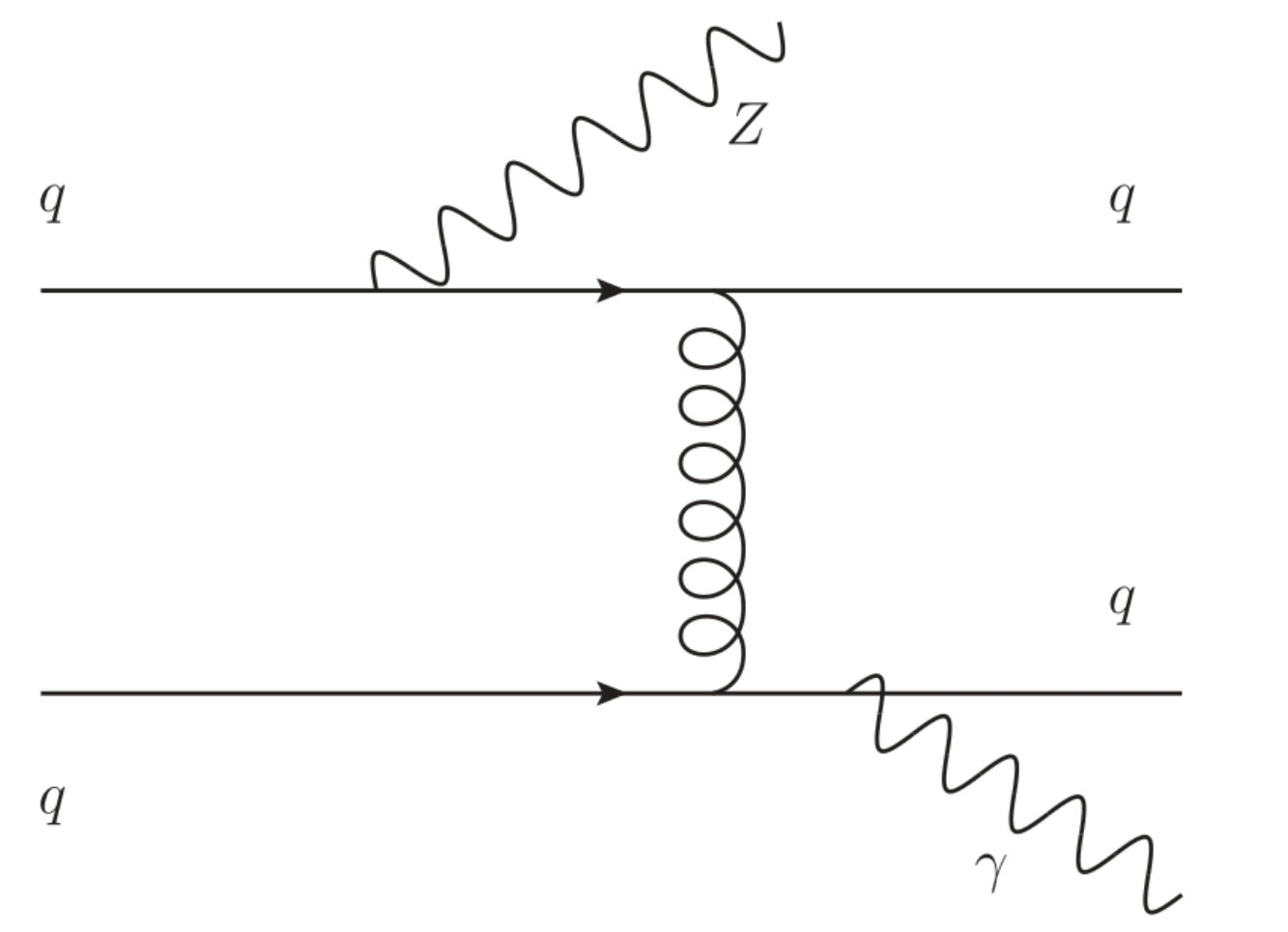} \hspace{1cm}
\includegraphics[width=0.33\columnwidth]{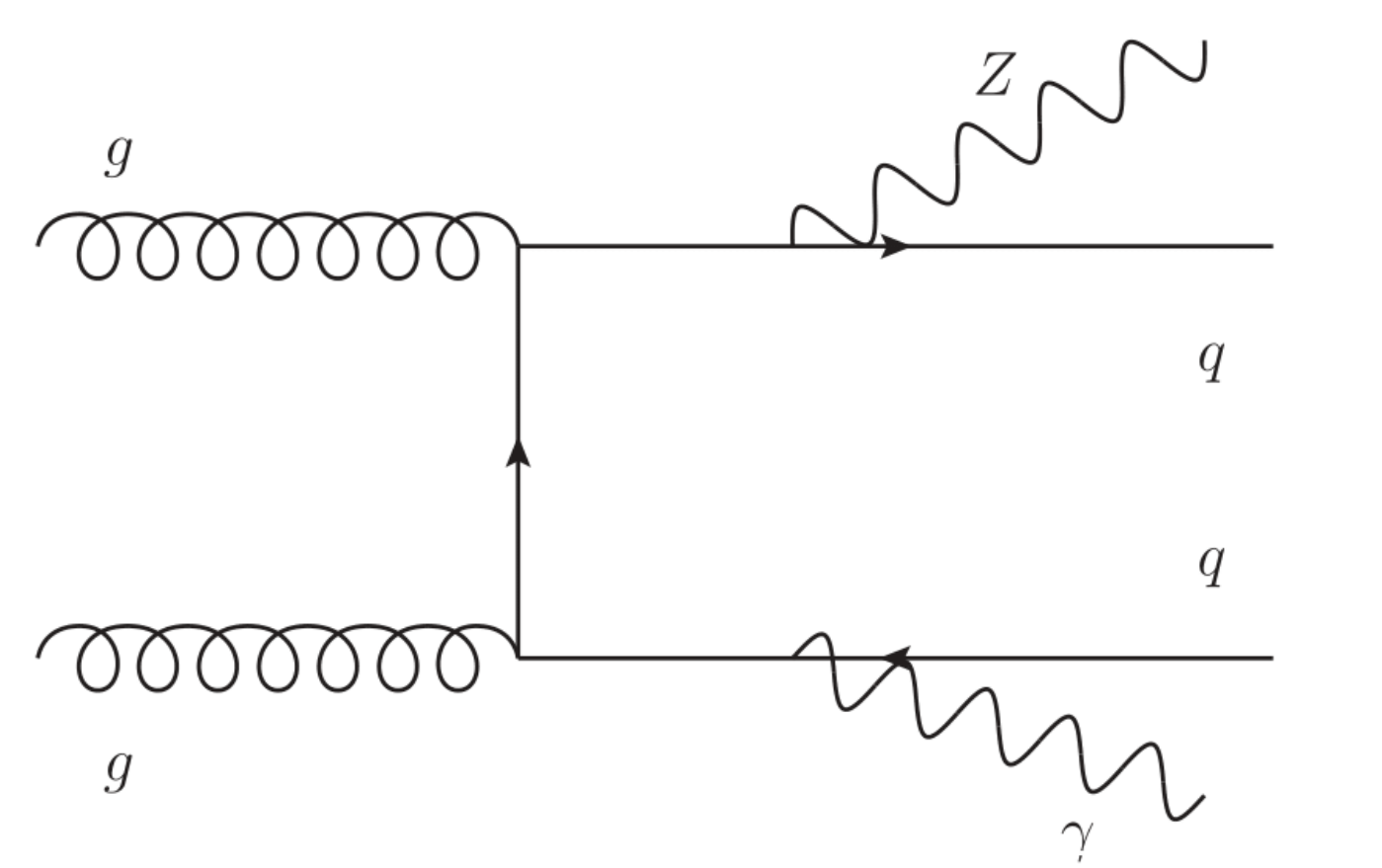}
\caption{Feynman diagrams of electroweak \ensuremath{Z \gamma jj}\xspace production involving the VBS subprocess (top left) or non-VBS subprocesses (top right) and of QCD \ensuremath{Z \gamma jj}\xspace production with gluon exchange (bottom left) or the s-channel $gg$--$qq$ process (bottom right).}
\label{fig:diagrams}
\end{center}
\end{figure}

The $Z\gamma jj$ final states are produced mainly through a combination of strong and electroweak interactions in proton--proton ($pp$) collisions.
The signal in this study is of the order $\alpha^4_\textsc{ewk}$ at tree level, where $\alpha_\textsc{ewk}$ is the electroweak coupling constant. In the following, such processes are referred to as \ensuremath{Z \gamma jj}\xspace EWK. The VBS process is an inseparable part of the gauge-invariant ensemble of \ensuremath{Z \gamma jj}\xspace electroweak processes~\cite{PhysRevD.74.073010}.
The main background in this study is of order $\alphas^2\alpha^2_\textsc{ewk}$ at tree level, where \alphas is the quantum chromodynamics (QCD) strong coupling constant. In the following, such processes are referred to as \ensuremath{Z \gamma jj}\xspace QCD. 
Example Feynman diagrams of the aforementioned processes are given in Figure~\ref{fig:diagrams}.
 
This paper presents a measurement of electroweak production in the \ensuremath{Z \gamma jj}\xspace final state, where the $Z$ boson decays into $\nu \bar{\nu}$. This choice is motivated by the fact that the $Z$ boson branching ratio into neutrinos is larger than the branching ratio into charged leptons; also, the background is under better control than in the hadronic decay channel.
The analysis uses $pp$ collision data recorded between 2015 and 2018 by the ATLAS detector~\cite{PERF-2007-01} during Run~2 at the LHC. 
Based on this measurement, a search for aQGCs is performed. Anomalous couplings produce deviations from the SM prediction that grow with increasing momentum transfer between the incoming partons. Hence, this analysis exploits a region of high momentum transfer by requiring a photon of transverse momentum larger than 150~\GeV.
 
The high energy and luminosity of the LHC allow rare VBS processes to be studied in detail.
In particular, the observation of the $Z(\ell\bar{\ell})\gamma jj$ channel has been reported~\cite{CMS-SMP-20-016}.
Moreover, the observation of \ensuremath{Z(\nu \bar{\nu}) \gamma jj}\xspace has been reported~\cite{EXOT-2021-17} in a low-energy phase-space region orthogonal to that presented in this paper. While the low-energy phase space gives negligible sensitivity for the aQGC search, it can be combined with the current high-energy analysis to increase the overall sensitivity to the SM electroweak \ensuremath{Z(\nu \bar{\nu}) \gamma jj}\xspace process.
Other $VVjj$ EWK processes, including $ZZjj$~\cite{STDM-2017-19}, $W\gamma jj$~\cite{CMS-SMP-19-008}, same-sign $WWjj$~\cite{CMS-SMP-17-004corr}, $W^{\pm}Zjj$~\cite{STDM-2017-23}, and photon-induced $W^+W^-jj$~\cite{STDM-2017-21}, have also been observed.


\section{Experimental set-up}
\label{sec:detector}

\newcommand{\AtlasCoordFootnote}{
A right-handed coordinate system is used with its origin at the nominal interaction point (IP) in the centre of the detector and the \(z\)-axis along the beam pipe. The \(x\)-axis points from the IP to the centre of the LHC ring, and the \(y\)-axis points upwards. Cylindrical coordinates \((r,\phi)\) are used in the transverse plane, \(\phi\) being the azimuthal angle around the \(z\)-axis. The pseudorapidity is defined in terms of the polar angle \(\theta\) as \(\eta = -\ln \tan(\theta/2)\). The angular distance between two physics objects is measured in units of \(\Delta R \equiv \sqrt{(\Delta\eta)^{2} + (\Delta\phi)^{2}}\).}

The ATLAS detector~\cite{PERF-2007-01} is a multi-purpose particle detector with a forward--backward symmetric cylindrical geometry and coverage of nearly the entire solid angle.\footnote{\AtlasCoordFootnote} It consists of an inner tracking detector (ID) surrounded by a thin superconducting solenoid providing a \SI{2}{\tesla} axial magnetic field, electromagnetic (ECAL) and hadronic (HCAL) calorimeters, and a muon spectrometer (MS).
 
The ID is used for precise measurements of charged-particle tracks. It is composed of two silicon detectors covering the pseudorapidity range \(|\eta| < 2.5\): a pixel detector (including the insertable B-layer~\cite{PIX-2018-001,ATLAS-TDR-19}) and a silicon microstrip tracker, surrounded by a straw-tube transition radiation tracker (TRT) with an acceptance of \(|\eta| < 2.0\), which also contributes to electron identification.
 
The ECAL is composed of high-granularity lead/liquid-argon (LAr) calorimeters in the region \(|\eta|< 3.2\) and copper/LAr calorimeters in the region \(3.2 < |\eta| < 4.9\). It plays a crucial role in photon identification, since photons are identified as narrow isolated showers in the ECAL. The HCAL consists of a steel/scintillator-tile calorimeter within \(|\eta| < 1.7\) and two copper/LAr and tungsten/LAr forward calorimeters within \(1.7 < |\eta| < 4.9\). The fine segmentation of the ATLAS calorimeter system allows efficient separation of jets from isolated prompt photons.
 
The MS comprises three large superconducting toroids, each having eight coils, as well as trigger and high-precision tracking chamber systems that cover the regions \(|\eta| < 2.4\) and \(|\eta| < 2.7\), respectively.
 
The ATLAS trigger system~\cite{TRIG-2019-04} has two levels, a hardware-based first-level trigger and a software-based high-level trigger (HLT). The trigger system selects events from the \SI{40}{\MHz} LHC proton bunch crossings at a rate of about \SI{1}{\kHz}.
 
An extensive software suite~\cite{ATL-SOFT-PUB-2021-001} is used in the reconstruction and analysis of real and simulated data, in detector operations, and in the trigger and data acquisition systems of the experiment.


\section{Data and simulation}
\label{sec:Data_and_MC}

The analysis uses the data collected by the ATLAS experiment from LHC $pp$ collisions at a centre-of-mass energy of $\sqrt{s}=13$~\TeV\ during 2015--2018 stable beam conditions, when all subdetectors were operational~\cite{DAPR-2018-01}, corresponding to a total integrated luminosity of 139~fb$^{-1}$~\cite{ATLAS-CONF-2019-021,LUCID2}.
 
Simulated signal and background events were produced with various Monte Carlo (MC) event generators, processed through the full ATLAS detector simulation~\cite{SOFT-2010-01}
using \textsc{Geant4}~\cite{Agostinelli:2002hh}, and then reconstructed with the same procedure as is used for data.
 
The effect of multiple $pp$ interactions in the same and neighbouring bunch crossings (referred to as \pileup) was modelled by overlaying the signal and background MC samples of simulated hard-scattering events with inelastic $pp$ events generated with \PYTHIA[8.186]~\cite{Sjostrand:2007gs}
using the \NNPDF[2.3lo] set of parton distribution functions (PDF)~\cite{Ball:2012cx} and the
A3 set of tuned parameters~\cite{ATL-PHYS-PUB-2016-017}.
 
The signal $Z(\nu\bar{\nu})\gamma jj$ EWK events, including samples with non-zero aQGC parameters as well as interference between $Z(\nu\bar{\nu})\gamma jj$ EWK and QCD production and background $W(\ell\nu)\gamma jj$ EWK and $t\bar{t}\gamma jj$ events, were generated using \MGNLO at leading order (LO) in both QCD and QED, interfaced to the \PYTHIA[8]~\cite{Sjostrand:2014zea} parton shower model. The decays of bottom and charm hadrons were simulated using the \EVTGEN~\cite{Lange:2001uf} program. Non-zero aQGC samples were generated for linear and quadratic BSM terms of the process amplitude for each effective field theory (EFT) operator considered.
 
For the signal $Z(\nu\bar{\nu})\gamma jj$ EWK sample, the next-to-leading-order (NLO) QCD corrections were produced with \VBFNLO[2.7.1]~\cite{VBFNLO}, taking into account the $m_{jj}$ dependence for consequent reweighting of \MGNLO result; the average value of the corrections is close to 1. The scale variations from \VBFNLO for the NLO QCD process are used instead of those from the LO \MGNLO sample.
EWK samples with an alternative parton shower model are obtained using \Herwig[7.13]~\cite{Bahr:2008pv,Bellm:2015jjp} instead of \PYTHIA[8] and used for the evaluation of that systematic uncertainty.

The $V\gamma$ and $V+$jets QCD backgrounds, prompt single-photon ($\gamma jj$) and multijet ($jj$) events, including $Z(\nu\bar{\nu})\gamma jj$, $W(\ell\nu)\gamma jj$, $Z(\ell\bar{\ell})\gamma jj$, $W(e\nu) jj$, $Z(\nu\bar{\nu}) jj$ and $Z(e\bar{e})$ production, were generated using the
\SHERPA~\cite{Bothmann:2019yzt} generator. Matrix elements at
NLO and LO QCD accuracy were matched and merged with
the \SHERPA parton shower based on Catani--Seymour
dipole factorisation~\cite{Gleisberg:2008fv,Schumann:2007mg} using the \MEPSatNLO
prescription~\cite{Hoeche:2011fd,Hoeche:2012yf,Catani:2001cc,Hoeche:2009rj}.
The virtual QCD corrections for the matrix elements at NLO accuracy were
provided by the \OPENLOOPS library~\cite{Buccioni:2019sur,Cascioli:2011va,Denner:2016kdg}. The samples using the \NNPDF[3.0nnlo]~\cite{Ball:2014uwa} PDF set were normalised to the next-to-next-to-leading-order (NNLO) prediction~\cite{Anastasiou:2003ds}.
 
An alternative $Z\gamma jj$ QCD sample was generated using \MGNLO[2.3.3]~\cite{Alwall:2014hca} at NLO interfaced to the \PYTHIA[8.212] parton shower model. The merging procedure in the event generation was performed using the FxFx scheme~\cite{Frederix:2012ps}.

Single-top or single-anti-top s-channel production was modelled using the \POWHEGBOX[v2]~\cite{Nason:2004rx,Frixione:2007vw,Alioli:2009je,Alioli:2010xd} generator, which provides matrix elements at NLO in the strong coupling constant in the five-flavour scheme with the \NNPDF[3.0nlo]~\cite{Ball:2014uwa} PDF set. The events were interfaced with \PYTHIA[8] using the \NNPDF[2.3lo] PDF set. The decays of bottom and charm hadrons were simulated using the \EVTGEN program.
 
The details of the matrix element generator, parton shower and parameter values (tune), PDF choice, and cross-section order for the processes mentioned above are listed in Table~\ref{tab:generators}.
 
\begin{table}[h]
\begin{center}
\caption{Simulated signal and background event samples used in the analysis with the corresponding matrix element and parton shower generators, cross-section order in perturbative QCD (pQCD) used to normalise the event yield, underlying-event tune and PDF set. Where indicated the NLO cross section is obtained with $K$-factors.}
\label{tab:generators}
\resizebox{0.98\linewidth}{!}{
\begin{tabular}{lllllll}
\toprule
Physics process& Generator& Parton shower  & Cross-section &Tune &PDF set\\
&				&					& order in pQCD  &	&			&			\\
\midrule
$Z(\nu\bar{\nu})\gamma jj$ EWK, $W(\ell\nu)\gamma jj$ EWK& \MGNLO[2.6.2]~\cite{Alwall:2011ujCorr}  &\PYTHIA[8.235] &LO, $K$-factor to NLO &A14~\cite{ATL-PHYS-PUB-2014-021} &\NNPDF[2.3lo] \\
\Zgjj interference, aQGC &\MGNLO[2.6.7]	&\PYTHIA[8.244]	& LO &A14 &\NNPDF[2.3lo]	\\
Alternative $Z(\nu\bar{\nu})\gamma jj$ EWK \& $W(\ell\nu)\gamma jj$ EWK& \MGNLO[2.6.2] &  \Herwig[7.13] &LO &A14 &\NNPDF[2.3lo] \\
$t\bar{t}\gamma jj$ &\MGNLO[2.3.3] & \PYTHIA[8.212] & LO, $K$-factor to NLO & A14 & \NNPDF[2.3lo] \\
Alternative $Z(\nu\bar{\nu})\gamma jj$ &\MGNLO[2.3.3] & \PYTHIA[8.212] & NLO & A14 & \NNPDF[2.3lo] \\
$Z(\nu\bar{\nu})\gamma jj$, $W(\ell\nu)\gamma jj$, $Z(\ell\bar{\ell})\gamma jj$ &\SHERPA[2.2.2] &\SHERPA[2.2.2] & NLO & default & \NNPDF[3.0nnlo]~\cite{Ball:2014uwa}\\
 
$W(e\nu)jj$, $W(\tau\nu)jj$, $Z(\nu\bar{\nu})jj$, $Z(e\bar{e})$  &\SHERPA[2.2.1] & \SHERPA[2.2.1] & NLO & default & \NNPDF[3.0nnlo] \\
$\gamma jj$ &\SHERPA[2.1] & \SHERPA[2.1]  & LO, $K$-factor to NLO & default & \CT[10nlo]~\cite{Lai:2010vv} \\
$jj$ &\SHERPA[2.1.1] & \SHERPA[2.1.1] & LO  & default & \CT[10nlo] \\
\multirow{2}{*}{$tjj$, $t\bar{t}jj$}	 &\multirow{2}{*}{\POWHEGBOX[v2]}& \multirow{2}{*}{\PYTHIA[8.230]} & \multirow{2}{*}{NLO} & \multirow{2}{*}{A14} & \NNPDF[3.0nlo],\\
& & & &  & \NNPDF[2.3lo]\\
\bottomrule
\end{tabular}
}
\end{center}
\end{table}


\section{Event reconstruction and selection}
\label{sec:selection}

Candidate $Z(\nu\bar{\nu})\gamma jj$ events are selected by requiring the presence of a highly energetic photon, high missing transverse momentum and two jets.
The lowest-threshold unprescaled single-photon trigger was chosen in order to select a high-energy phase-space region, which is sensitive to aQGC.
This trigger requires a transverse energy $\et^{\gamma} > 140$~\GeV\ and applies a `loose' photon identification criterion~\cite{PERF-2017-02}. The trigger efficiency for the photon candidates reconstructed offline and passing the `tight' identification selection is more than 98.5\%~\cite{CERN-EP-2019-169}.
 
\subsection{Object reconstruction}
\label{sec:reconstruction}
 
Photons are reconstructed~\cite{EGAM-2018-01} from clusters of energy deposited in the ECAL and selected to pass $|\eta| <$~2.37 and $\et^{\gamma} > 150$~\GeV\ requirements. Clusters that are matched to one or two tracks originating from a conversion vertex are classified as converted photon candidates, whereas clusters without a matching track or reconstructed conversion vertex in the ID are classified as unconverted photon candidates. Electron candidates are reconstructed~\cite{EGAM-2018-01} from ECAL energy clusters and matched to a track reconstructed in the ID. They are required to have $|\eta| <$~2.47 and transverse momentum $\pt > 7$~\GeV. Both the photon and electron candidates must be outside of the calorimeter barrel/endcap transition region ($1.37< |\eta| <1.52$).
 
Muons are reconstructed~\cite{MUON-2018-03} from tracks in the MS matched to a corresponding track in the ID (referred to as `combined muons'). The combined track is required to have $\pt > 7$~\GeV\ and $|\eta| <$~2.7. Electron and muon tracks are required to originate from the primary vertex.\footnote{
The primary vertex is identified as the vertex with the highest scalar sum of the squared transverse momenta in the event.} The transverse impact parameter significance\footnote{
The transverse impact parameter significance is defined as $|d_0|/\sigma(d_0)$, where $d_0$ is the distance of closest approach of $e$ or $\mu$ to the primary vertex in the transverse plane with an uncertainty $\sigma(d_0)$.} is required to be less than 5 and 3, respectively, for electrons and muons. The longitudinal impact parameter\footnote{
The longitudinal impact parameter is equal to $|z_0 \cdot \sin\theta|$, where $z_0$ is the difference between the value of the $z$ coordinate of the point on the track at which $d_0$ is defined and the longitudinal position of the primary vertex.} must be less than 0.5~mm for both the electrons and muons.
 
Photons and electrons are required to meet identification criteria based on their shower shapes in the ECAL, the amount of energy leaked into the hadronic calorimeter, and ID tracking information. The `tight' photon identification criterion corresponding to passing all the requirements on shower shape variables is used in the analysis. The identification efficiency for the `tight' photons is greater than 88\% in Run~2~\cite{EGAM-2018-01}. `Loose' photons are selected in order to model jets misidentified as photons for the data-driven background estimation methods described in \Sect{\ref{sec:bkg_est}}. Electron candidates are required to satisfy the `loose' electron identification criterion~\cite{EGAM-2018-01}. The `medium' identification requirement is used for muon candidates~\cite{MUON-2018-03}.
 
Photons must satisfy the `tight' isolation criterion~\cite{EGAM-2018-01}, thereby fulfilling two requirements. Firstly, the sum of the transverse energies (at the electromagnetic energy scale) of positive-energy topological clusters located within a distance $\Delta R = 0.4$ of the photon candidate must be less than $0.022 \cdot \et^\gamma + 2.45$~\GeV. Secondly, the scalar sum of the transverse momenta of the tracks located within a distance $\Delta R = 0.2$ of the photon candidate must be less than $0.05 \cdot \et^\gamma$. The `loose' isolation requirement is imposed on electrons and muons~\cite{EGAM-2018-01, MUON-2018-03}.
 
To suppress the beam-induced background~\cite{DAPR-2012-01}, the $z$-axis coordinate pointed to by the photon candidate is required to be less than 250~mm from the identified primary vertex.
 
Jets are reconstructed from topological clusters in the calorimeters using the anti-$k_t$ algorithm~\cite{Cacciari:2008gp,Fastjet} with a radius parameter of $R = 0.4$ and are required to have $\pt > 20$~\GeV. Jets are fully calibrated using the jet energy scale derived from 13~\TeV\ data and simulation~\cite{PERF-2016-04} and corrected for pile-up effects~\cite{ATL-PHYS-PUB-2015-015}. To suppress jets originating from pile-up, jets with $|\eta| < 2.5$ and $20 < p_\text{T} < 120$~GeV must satisfy the `medium' identification requirement placed on the jet vertex tagger (JVT) output~\cite{PERF-2014-03}. The forward pile-up jet vertex tagger (fJVT)~\cite{PERF-2016-06}, which can be applied to jets with $2.5 < |\eta| < 4.5$ and $20 < p_\text{T} < 120$~GeV, causes marginal changes, so it is not performed. The selected jets are required to have $\pt > 50$~\GeV\ and $|\eta| <$~4.5.
 
The missing transverse momentum $\metpt$ is computed as the negative vector sum of the transverse momenta of candidate leptons with $\pt > 7$~\GeV, photons with $\pt > 10$~\GeV, jets with $\pt > 20$~\GeV, and tracks from the primary vertex not associated with any physics objects (the `soft term')~\cite{PERF-2016-07}. The quantity \met is defined as the magnitude of $\metpt$ and is used as a measure of the total transverse momentum of particles not registered by the detector.
 
Possible double counting of contributions from reconstructed particles is avoided by applying an ambiguity resolution procedure. The objects are removed in the following order: first, an electron lying within a $\Delta R = 0.1$ cone around a muon, then a selected jet without any JVT requirement lying within $\Delta R = 0.3$ of a photon, muon, or electron, and, finally, a photon lying within $\Delta R = 0.4$ of either a muon or an electron.
 
\subsection{Region definitions}
\label{sec:reg_def}
 
The $Z\gamma$ inclusive region requires events with exactly one `tight' isolated photon with transverse energy $E_\mathrm{T}^{\gamma} >150$~\GeV\ and at least two jets. Selected events must also have $\met > 120$~\GeV\ and calorimeter-measured $\met$ significance\footnote{
$\met$ significance is calculated as $|\vec{p}_{\mathrm{T}}^{\mathrm{\;miss}}|^2/\left(\sigma_\text{L}^2\left(1-\rho_\text{LT}^2 \right) \right)$, where $\sigma_\text{L}$ is the total variance in the direction longitudinal to the $\met$, and $\rho_\text{LT}$ is the correlation coefficient of the longitudinal (L) and transverse (T) measurements~\cite{ATLAS-CONF-2018-038}.} greater than 12. An angular separation between both the leading (highest-\pT) and subleading (second-highest-\pt) jets $j_\text{1,2}$ and $\metpt$ of $|\Delta\phi(j_\text{1/2}, \metpt)| > 0.3$ is also required. These criteria reduce the number of background events characterised by \met arising from instrumental sources or poorly reconstructed physics objects (fake \met), mainly due to the $\gamma jj$ process. A lepton veto, which discards events with any electron or muon candidate, is applied to reduce the contribution from $W(\ell\nu)\gamma jj$ and $Z(\ell\bar{\ell})\gamma jj$ backgrounds. An azimuthal angle requirement $|\Delta\phi(\gamma, \vec{p}_{\mathrm{T}}^{\mathrm{\;miss}})| > 0.4$ significantly decreases the $W(e\nu)jj$ background. Finally, a restriction on the soft term of $\metpt$ reconstructed from ID tracks, $p_\mathrm{T}^{\text{SoftTerm}} < 16$~\GeV, is effective against most backgrounds.
 
The $W\gamma$ control region (CR) requires events with at least one lepton. All other selection criteria are the same as for the $Z\gamma$ inclusive region.
 
\begin{figure}
\centering
\includegraphics[width=0.8\linewidth]{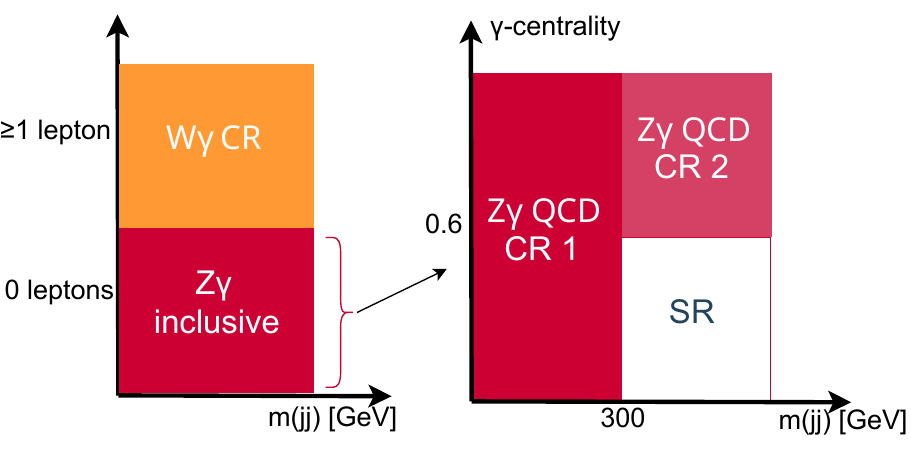}
\caption{Definition of the regions used in the analysis.}
\label{fig:regionDefinition}
\end{figure}

The $Z\gamma$ inclusive region is divided into three subregions as shown in~\Fig{\ref{fig:regionDefinition}}. The signal region (SR) is required to have $\mjj > 300$~\GeV\ and $\gamma$-centrality\footnote{
Photon centrality relative to the two jets with the highest \pt values in the event is defined as $\gamma$-centrality = $\left|\frac{y(\gamma) - 0.5 [y(j_1)+y(j_2)]}{y(j_1)-y(j_2)}\right|$, where $y = 0.5 \times \text{ln}[(E + p_z)/(E - p_z)]$ is the rapidity of the objects ($p_z$ is the $z$-component of the momentum of a particle).}
< 0.6, where $\mjj$ is defined as the invariant mass of the two jets with the highest values of \pt in the event. The $Z\gamma$ QCD CR\,1 requires events with $\mjj < 300$~\GeV; it is used to estimate the $Z(\nu\bar{\nu})\gamma jj$ QCD background yield. The $Z\gamma$ QCD CR\,2 has the same selection criteria as the SR but requires events with $\gamma$-centrality > 0.6; it is used to check for possible \mjj~mismodelling. The values of the requirements are chosen to maximise the number of events and the purity of the targeted process in each region.
 
\section{Background estimation}
\label{sec:bkg_est}

The main background for the \ensuremath{Z \gamma jj}\xspace EWK process is QCD production of $Z(\nu\bar{\nu})\gamma jj$. This background constitutes 36\% of the total predicted event yield in the SR.\footnote{The indicated percentages of background contributions are calculated using the predicted event yield after the fit described in Section~\ref{sec:sig_extraction}.} It is estimated from a simultaneous fit of the MC distributions to data in the control regions described in Section~\ref{sec:selection}. The fit procedure is detailed in Section~\ref{sec:sig_extraction}.
 
Other well-modelled backgrounds~\cite{STDM-2011-17,CMS-SMP-19-008,TOPQ-2020-03} arise from $W(\ell\nu)\gamma jj$ QCD and EWK production, and production of $t\bar{t}\gamma jj$ with semileptonic or fully leptonic decays. They contribute 25\%, 7\% and 6\% to the total predicted event yield in the SR repectively. Their distribution shapes are taken from MC predictions and their normalisation is obtained from the simultaneous fit. Varying the ratio of $W(\ell\nu)\gamma jj$ to $t\bar{t}\gamma jj$ contributions has negligible impact on the fit result.
 
The minor background from $Z(\ell\bar{\ell})\gamma jj$ production is estimated from MC simulation and without normalisation via the fit (it is less than 1\% of the predicted event yield in the SR). In all these processes except $Z(\nu\bar{\nu})\gamma jj$, the leptons are either not reconstructed or they are $\tau$-leptons which decay into hadrons.
 
Background processes with object misidentification or incorrect energy measurement are not well modelled by the MC simulation, and so they are estimated from data. There are three such background processes: \etogam misidentification, \ETmiss mismeasurement, and \jtogam misidentification. They respectively contribute 6\%, 5.5\%  and 2\% to the total predicted event yield in the SR. These backgrounds are included in the simultaneous fit, with the normalisation estimated from data.
 
The sources of the \etogam misidentification background are mainly processes with a $W$ boson decaying leptonically, which are $W(e\nu)jj$, $tjj$, and $t\bar{t}jj$ production. At high energies, such misidentification mainly occurs when a prompt electron is mistaken for a prompt photon, e.g. if an electron's track is not reconstructed in the ID. The \etogam misidentification rate ($f_{\etogam}$) is estimated from data using a variation of the tag-and-probe method (e.g.\ in Ref.~\cite{Straessner:1354502}), in which it is assumed that $e\gamma$ pairs with invariant mass near the $Z$ boson mass contain an electron or positron misidentified as a photon. The probe photon is selected in the same way as the SR photon, while the probe electron is selected with a method that is as close as possible to the selection of the SR photon. In this case, $f_{\etogam}$ can be estimated as the ratio of the number of events containing $e\gamma$ to the number containing $e^+e^-$ tag-and-probe pairs in the $Z$-peak region after subtraction of the Drell--Yan background. The subtraction is performed using extrapolation of an exponential polynomial fit of the Drell--Yan background outside the $Z$-peak region. The measured $f_{\etogam}$ ranges from 2\% to 6.8\% depending on photon $\eta$ and $\ET$. The systematic uncertainty of the $f_{\etogam}$ estimate ranges from 3.7\% to 16\%. It has three contributions, which are, listed in decreasing order of magnitude, the uncertainty from a check of the method's validity in $Z(ee)$ simulation, the uncertainty from the fit to estimate the background under the $Z$ peak, and the choice of $Z$-peak region. The statistical uncertainty of $f_{\etogam}$ ranges from 2.5\% to 6.3\%. The measured value of $f_{\etogam}$ is then used to estimate the background yield. The corresponding $W(e\nu)jj$-enriched regions are built in data for the $W\gamma$ CR, the $Z\gamma$ inclusive region, the $Z\gamma$ QCD CR\,1 and CR\,2, and the SR, with a probe electron selected instead of a photon. The resulting event yields or distributions are multiplied by $f_{\etogam}$, taking into account $\eta$ and $\ET$ dependencies. The total systematic uncertainty of the background estimate includes systematic and statistical uncertainties of the $f_{\etogam}$ estimate and also the impurity of the probe-electron control region. According to MC simulations, the contamination consists of events with fake electrons from $j\rightarrow e$ misidentification and varies from less than 1\% to 2\% depending on the region; these values are taken as the systematic uncertainties.
The total systematic uncertainty of the \etogam background estimate varies from 4.7\% to 7.4\%.
 
The mismeasured \ETmiss in \gj production occurs when there is an incorrect measurement of the jet energy or when some of the jets are not reconstructed in the event. To estimate this background, a two-dimensional sideband (ABCD) method (e.g.\ see Ref.~\cite{STDM-2010-08}) based on the \ETmiss significance and $p_\mathrm{T}^{\text{SoftTerm}}$ discriminating variables is used. Region A corresponds to the $Z\gamma$ inclusive region. Orthogonal control regions B, C, and D are built by inverting either of the analysis selections on \ETmiss significance or $p_\mathrm{T}^{\text{SoftTerm}}$, or both. The discriminating variables are chosen to ensure that the correlation between them is small and to minimise leakage of signal events into the control regions. In this case, the relation between the numbers of background events in the ABCD regions is $N_\text{A}/N_\text{B}=N_\text{C}/N_\text{D}$. The background of \gj in the $Z\gamma$ inclusive region is estimated using the yields observed in the B, C, and D regions in the data, where contamination with non-\gj events is removed using either MC simulation or data-driven estimation.
The correlation factor\footnote{The correlation factor in MC is determined as $R = \frac{N_A\cdot N_D}{N_B\cdot N_C}$, where $N_i$ is the number of background events in the corresponfing control region. In absence of correlation $R=1$.} is estimated from \gj MC events and is $1.09\pm0.18$; a value of 1 would indicate an absence of correlation.
The statistical uncertainty of the fake-\ETmiss background is assessed by independently varying the non-\gj backgrounds in the control regions by $\pm1\sigma$. It results in a 44\% uncertainty in this background estimate for the $Z\gamma$ inclusive region. The total systematic uncertainty is 32\%. The dominant contribution (31\%) is
obtained by varying the correlation factor by its uncertainty.
To obtain estimates of this background in the signal and control regions of the analysis (the $Z\gamma$ QCD  CRs and $W\gamma$ CR), the distributions of $m_{jj}$ and $\gamma$-centrality from the \gj MC events are used. They are found to be in agreement with data  within the uncertainties, after subtraction of other backgrounds. The \gj MC events are used to model the shape of this background in the fit for the observable used for the cross-section extraction.
 
Background from \jtogam misidentification arises mainly from $Z$ boson production where the $Z$ boson decays into a neutrino--antineutrino pair and from multijet production combined with fake \ETmiss. This background is also estimated using the ABCD method, with A region corresponding to the $Z\gamma$ inclusive region. The discriminating variables are those used in the photon isolation and ECAL shower-shape identification criteria. To construct the orthogonal regions, the SR requirement for the photon to satisfy the `tight' identification criteria is replaced by a `non-tight' requirement for the C and D regions, and the ECAL-based isolation criterion is inverted in the B and D regions. The `non-tight' photon candidate must fail at least one of the two shower shape selections~\cite{PERF-2013-04} $w_\text{s3}$\footnote{Front lateral width (3 strips) measures the shower width along $\eta$ in the first layer of the electromagnetic calorimeter, using a total of three strip cells centred on the largest energy deposit. It is calculated as $w_\text{s3}=\sqrt{\frac{\Sigma E_i(i-i_\text{max})^2}{\Sigma E_i}}$. The index $i$ is the strip identification number, $i_\text{max}$ identifies the strip cell with the most energy, and $E_i$ is the energy deposit in each strip cell.} and $F_\text{side}$\footnote{Front side energy ratio measures the lateral containment of the shower along the $\eta$ direction. It is calculated as $F_\text{side}=\frac{E(\pm3)-E(\pm1)}{E(\pm1)}$, where $E(\pm n)$ is the energy in
the $\pm n$ strip cells around the one with the most energy.} which are associated with the strip layer of the ECAL. These are chosen instead of other `non-tight' variables because of their lower correlation with isolation variables used in the construction of the regions. In addition, the correlation factor obtained from $Z(\nu\bar{\nu})jj$ inclusive MC simulation and the one obtained from data agree best when using this `non-tight' definition. The statistical uncertainty of the \jtogam background is obtained in the same way as for the \gj background.  
The largest uncertainty in the background estimate is 52\% for the $Z\gamma$ inclusive region.
The dominant systematic uncertainty comes from variations of the control region definition~\footnote{Variations consist in the choice of non-tight identification criteria and the choice of the energy gap between regions with normal and inverted isolation criteria} and is 18\%, while the total is 19\%.
The $Z(\nu\bar{\nu})\gamma jj$ QCD MC sample is used as a template for the background in the fit and for the extrapolation to the signal and control regions of the analysis (these are the $Z\gamma$ QCD control regions and $W\gamma$ control region). The reason why the $Z(\nu\bar{\nu})\gamma jj$ QCD MC sample is used instead of the $Z(\nu\bar{\nu})jj$ inclusive MC sample is that far fewer events are selected from the latter; the distributions from these MC samples agree within their uncertainties.
 
Another background, due to \pileup, arises when the photon and $Z$ boson are produced in different $pp$ collisions in the same LHC bunch crossing. This background is estimated from the distribution of the longitudinal separation between the reconstructed primary vertex and the reconstructed coordinate of a photon's origin. Only converted photons are used since they have better $z$-coordinate resolution. This contribution has no significant effect on the overall background shape and is considered only as a systematic normalisation uncertainty of 1.9\% in the fit.


\section{Systematic uncertainties}
\label{sec:syst}

Experimental sources of systematic uncertainty include uncertainties in the energy scale and resolution of jets, photons and electrons, in the scale and resolution of the muon momentum, and in the missing transverse momentum.
Additional contributors to the experimental uncertainty are the uncertainties in the scale factors used to reproduce the trigger, reconstruction, identification and isolation efficiencies, and pile-up conditions measured in data.
The uncertainty in the combined 2015--2018 integrated luminosity is 1.7\%~\cite{ATLAS-CONF-2019-021}, obtained using the LUCID-2 detector~\cite{LUCID2} for the primary luminosity measurements.
 
The theoretical systematic uncertainties considered in this analysis are related to higher-order QCD corrections and our choice of PDF and value of the strong coupling constant $\alphas$.
The uncertainties due to higher-order QCD corrections are estimated by varying the renormalisation and factorisation scales by factors of one-half and two, and ignoring the combinations that differ by a factor of four.
The uncertainties due to the PDF and $\alphas$ choice are estimated using the PDF4LHC prescription~\cite{Butterworth:2015oua}.
Additionally, a global modelling uncertainty was evaluated using an alternative MC generator either for the matrix element generation, underlying event and parton showering or for only the last two of these.
The alternative PDF sets and MC generators are described in \Sect{\ref{sec:Data_and_MC}}.
 
\begin{figure}
\centering
\subfloat[]{
\includegraphics[width=0.49\linewidth]{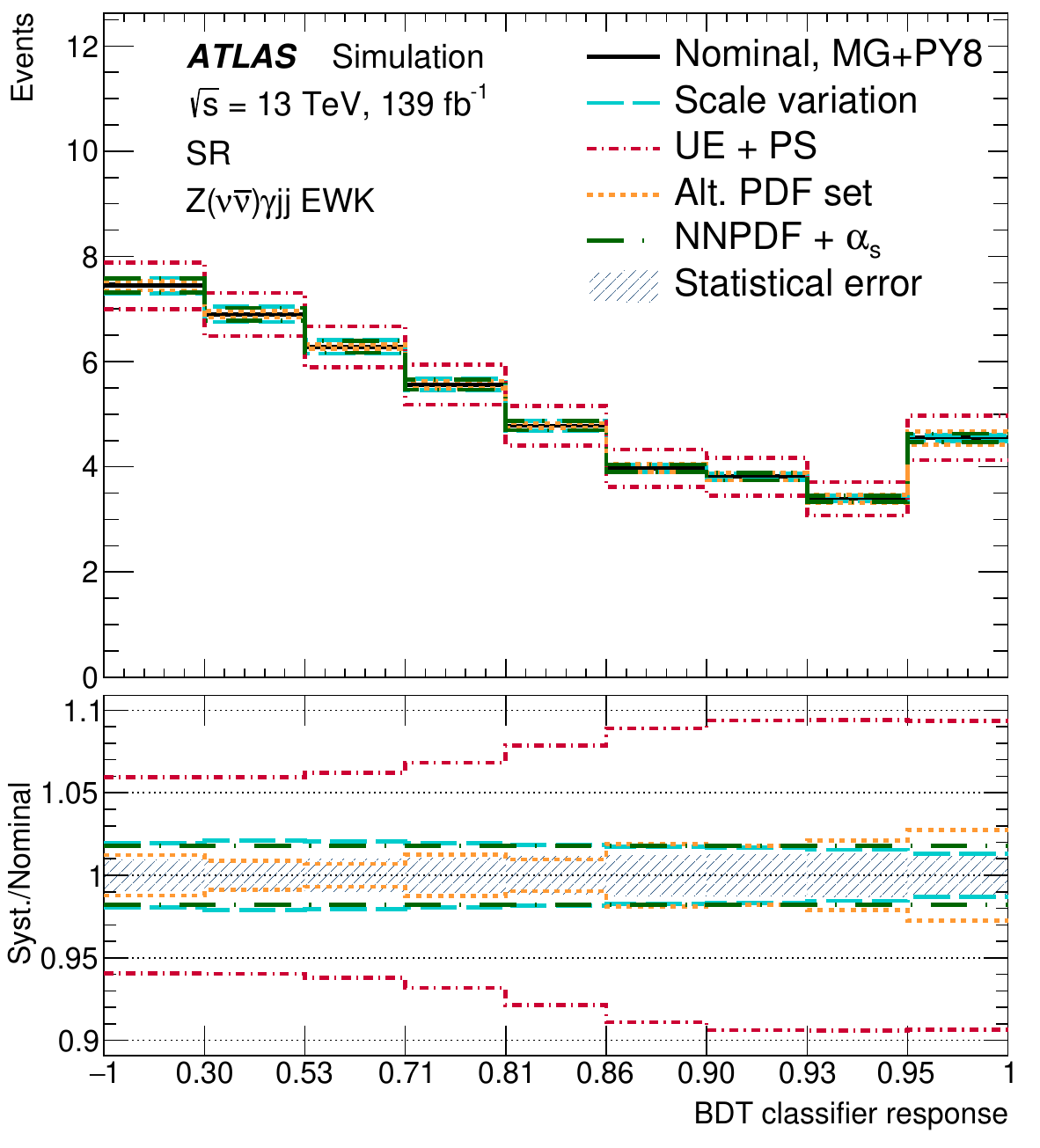}
\label{fig:theorySystEffectZgEWK}
}
\subfloat[]{
\includegraphics[width=0.49\linewidth]{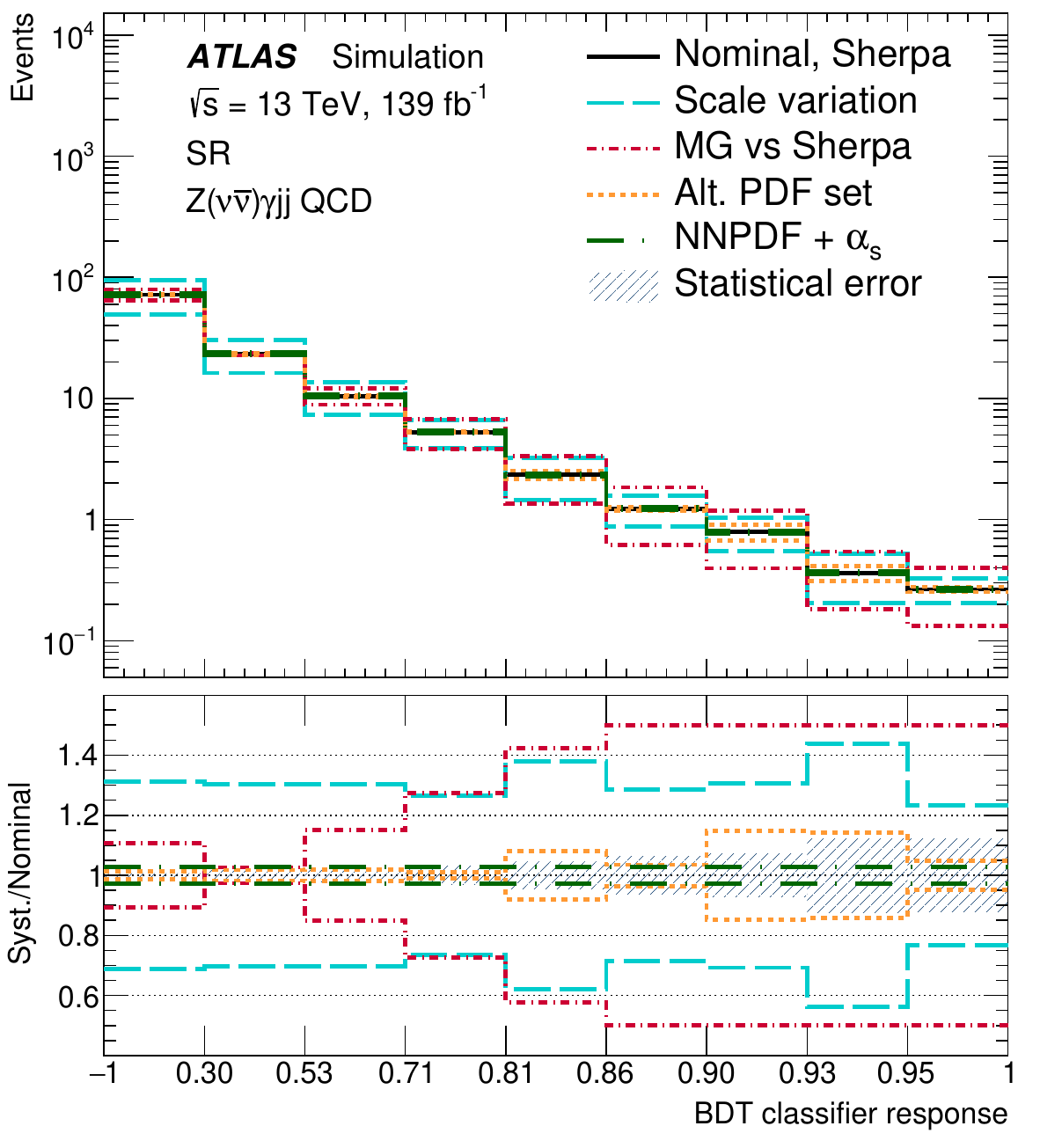}
\label{fig:theorySystEffectZgQCD}
}
 
\caption{These graphs indicate the effect of the main theory uncertainties, which are associated with the renormalisation and factorisation scales (dashed cyan), underlying event and parton showering (UE+PS) or generator choice (dash-dotted red), alternative PDF sets (dotted orange), combined NNPDF set variation and $\alphas$ uncertainty (loosely dash-dotted green). These are shown in the signal region for \protect\subref{fig:theorySystEffectZgEWK} the $Z(\nu\bar{\nu})\gamma jj$ EWK process and \protect\subref{fig:theorySystEffectZgQCD} the $Z(\nu\bar{\nu})\gamma jj$~QCD process. The BDT classifier response was remapped into equal width bins for better representation. The uncertainty band corresponds to the uncertainty due to the limited number of MC events.}
\label{fig:theorySystEffect}
\end{figure}

The signal selection uses a boosted decision tree (BDT). The signal-to-background ratio is expected to increase at high values of the BDT classifier
as is described in Section~\ref{sec:sig_extraction}.
The effect of the theoretical systematic uncertainties on the $Z(\nu\bar{\nu})\gamma jj$ EWK and $Z(\nu\bar{\nu})\gamma jj$ QCD processes, versus BDT classifier response, is shown in \Fig{\ref{fig:theorySystEffect}}.
 
The yield of the interference between the $Z(\nu\bar{\nu})\gamma jj$ EWK signal and QCD background is estimated to be small (5.8\% of the total signal yield in the SR) so it is not included as a part of the electroweak signal in the fit.
Instead, the signal is represented by the pure electroweak process, and the directly generated interference contribution is taken as an extra signal uncertainty.
 
Additional uncertainties related to data-driven background estimates are also considered.
Systematic uncertainties are assigned to the normalisation of the backgrounds from \etogam{} misidentification, \jtogam{} misidentification, incorrect \met{} measurement, and the combination of a $Z$ boson and a photon from different $pp$ collisions (pile-up background), with values corresponding to those described in \Sect{\ref{sec:bkg_est}}.

Mismodelling of the \mjj{} distribution is observed in the $Z\gamma$ QCD CR\,2 (see \Fig{\ref{fig:observedPostFitZgQCD_2}}, where the prediction-to-data ratio has $\chi^2/N_{\textrm{df}} = 8.2 / 7$ and a $p\textrm{-value} = 0.31$) and could have a similar impact in the SR.
Since the $Z(\nu\bar{\nu})\gamma jj$~QCD and $W(\ell\nu)\gamma jj$~QCD  processes are the dominant ones in this region (contributing 45\% and 31\% of the total predicted event yield, respectively) and the $W(\ell\nu)\gamma jj$~QCD process is shown to be accurately modelled in the $W\gamma$ CR (see \Fig{\ref{fig:observedPostFitWg}}), the mismodelling is attributed to the $Z(\nu\bar{\nu})\gamma jj$~QCD process.
Due to the similarities between these two processes (being produced by the same MC generator and having final-state kinematical properties that differ only because of the final-state boson) and the fact that the $W\gamma$ CR has data in all of the \mjj{} and $\gamma\textrm{-centrality}$ spectra bins, the $Z(\nu\bar{\nu})\gamma jj$~QCD modelling in the SR was checked with the $W(\ell\nu)\gamma jj$~QCD process in the analogous part of the $W\gamma$ CR.
This validates the modelling of $Z(\nu\bar{\nu})\gamma jj$~QCD in the SR to the level of agreement between the $Z(\nu\bar{\nu})\gamma jj$~QCD and $W(\ell\nu)\gamma jj$~QCD processes, and to the level of agreement between the  $W(\ell\nu)\gamma jj$~QCD process and the data.
The discrepancies are used to assign two systematic uncertainties to the $Z(\nu\bar{\nu})\gamma jj$~QCD contribution in the SR, resulting in 15\% and 23\% relative uncertainties in the total event yield, respectively.


\section{Signal extraction procedure and results}
\label{sec:sig_extraction}

A BDT classifier created with the TMVA~\cite{TMVA} package is used to separate the signal from the background processes described in \Sect{\ref{sec:bkg_est}}.
It is trained in the $Z\gamma$ inclusive region.
Due to the low number of events in the $Z\gamma$ inclusive region, the \gj, $Z(\ell\bar{\ell})\gamma$ and $Zj$, and multijet ($jj$) backgrounds are not used to train the classifier.
In order to utilise the full set of simulated events, two classifiers are created and trained on two statistically independent samples.
The classifier trained on one sample is applied to the other, and the final response distribution is built by combining the responses of these two classifiers.
The following variables are used to create the classifier:
$m_{jj}$,
$\Delta y(j_1, j_2)$,
\met,
$\pT$-balance,\footnote{The $\pT\textrm{-balance} = \frac{|\metpt + \vec{p}_{\mathrm{T}}^{\,\gamma} + \vec{p}_{\mathrm{T}}^{\,j_1} + \vec{p}_{\mathrm{T}}^{\,j_2}|}{\met + \ET^{\gamma} + \pT^{j_1} + \pT^{j_2}}$.}
$\eta(j_2)$,
$\pT(j_1)$,
$\eta(\gamma)$,
$\pT$-balance (reduced),\footnote{The $\pT\textrm{-balance (reduced)} = \frac{|\vec{p}_{\mathrm{T}}^{\,\gamma}+\vec{p}_{\mathrm{T}}^{\,j_1}+\vec{p}_{\mathrm{T}}^{\,j_2}|}{\ET^{\gamma}+\pT^{j_1}+\pT^{j_2}}$.}
$N_{\text{jets}}$,
$\sin \left(\left| \Delta \varphi(j_1, j_2)/2 \right|\right)$ and
$\Delta y(j_1, \gamma)$.
This is the smallest set of variables that gives the classifier sufficient separation power.
Although the $\gamma$-centrality provides as much separation power as \mjj{} or $\Delta y(j_1, j_2)$, it was not used in the classifier because this allows the creation of an independent $Z\gamma$ QCD CR\,2 that was used to check the modelling of the \mjj{} distribution.
 
To extract the $Z(\nu\bar{\nu})\gamma jj$~EWK cross section, a binned maximum-likelihood fit~\cite{Cranmer:1456844} is performed using the BDT classifier response distribution in the SR and the \mjj{} distributions in the $Z\gamma$ QCD CRs~1 and~2 and the $W\gamma$ CR; this combination uses 31 bins.
Three free parameters are introduced in the combined fit: a signal strength parameter, $\mu_{Z\gamma\text{EWK}}$, and two normalisation factors for the main background sources.
The first, $\mu_{Z\gamma\text{QCD}}$, is used to scale the $Z(\nu\bar{\nu})\gamma jj$~QCD process yield, while $\mu_{W\gamma}$ is used to scale the yields of the $W(\ell\nu)\gamma jj$~QCD, $W(\ell\nu)\gamma jj$~EWK and $t\bar{t}\gamma jj$ processes because of their similar final states.
 
Each source of systematic uncertainty is implemented in the likelihood function as a nuisance parameter (NP) with a Gaussian constraint.
All instrumental uncertainties, as well as the pile-up background uncertainty, are treated as correlated between the processes and regions.
Almost all of the other uncertainties are treated as correlated between regions. The exceptions are the scale uncertainties for all of the processes and the modelling uncertainties of the $Z(\nu\bar{\nu})\gamma jj$~QCD and $W(\ell\nu)\gamma jj$~QCD processes.
All scale uncertainties are uncorrelated between the regions.
The modelling uncertainties are separated into two groups: one correlated between all of the CRs and one for the SR.
This conservative approach avoids unnecessary constraints upon systematic uncertainties coming from the differences in the MC generators.
 
To account for the effect of the limited size of the simulated samples, an uncertainty with a Poissonian constraint is introduced for bins with a relative statistical uncertainty higher than 5\%.
The signal strength, background normalisation coefficients and yields for all of the processes are estimated in the fit to the observed data in the signal and control regions.
 
The observed significance is estimated by setting $\mu_{Z\gamma\text{EWK}}=0$ and performing a background-only fit to the data in all of the regions so as  to determine the probability of rejecting the background-only hypothesis.
In this approach uncertainties that only affect the signal process (i.e. theoretical uncertainties and the $Z(\nu\bar{\nu})\gamma jj$ EWK/QCD interference) have no effect on the significance calculation.
The expected significance is estimated by fitting the artificial Asimov dataset in the same way.
Such a dataset is constructed by modifying the predicted values with the normalisation coefficients and NPs obtained in the fit in the CRs while assuming no signal is present~\cite{Cowan:2010js}.
 
The signal strength is measured to be:
\begin{equation*}
\begin{aligned}
\mu_{Z\gamma\text{EWK}} 
& = 0.78^{+0.25}_{-0.23}\text{ (stat.)} ^{+0.21}_{-0.17}\text{ (syst.)}.
\end{aligned}
\end{equation*}
 
The observed (expected) significance of the result is $3.2\sigma$ $(3.7\sigma)$.
The $\mu_{Z\gamma\text{QCD}}$ and $\mu_{W\gamma}$ normalisation coefficients are measured to be $1.21^{+0.37}_{-0.31}$ and $1.02^{+0.22}_{-0.17}$, respectively, signifying agreement with the predicted yields within the uncertainties.
 
The predicted fiducial cross section is computed in the phase space defined in \Tab{\ref{table:FidVol}}.
The definition of the fiducial phase space closely follows the detector-level selections, using photons, electrons, muons, $\met$ and jets at the particle level. These stable final-state particles (with proper decay length $c\tau > 10$~mm) are produced in the hard scatter; this includes those that are the products of hadronisation. Thus they are reconstructed in simulation, prior to their interactions with the detector.
The leptons used in the veto are reconstructed at the particle level, with a correction for fully recovered final-state radiation applied.
No requirement is placed on the $\met$ significance or $p_\mathrm{T}^{\text{SoftTerm}}$ due to the complexity of defining these variables at particle level; however, the detector-level $\met$ requirement is applied to the particle-level $\met$, which corresponds to the $\et$ of the dineutrino system. All the other kinematic selection requirements are the same as those at detector level in Section~\ref{sec:reg_def}. The fiducial region selection efficiency is 33\%.
The fiducial cross section was predicted with \MGNLO (interfaced with \PYTHIA) at  leading order, with next-to-leading-order QCD corrections and scale uncertainties computed with \VBFNLO. Its value is
\begin{equation*}
\sigma_{Z\gamma\text{EWK}}^{\text{pred}}
= 0.98 \pm 0.02 \text{ (stat.)} \pm 0.09 \text{ (scale)} \pm 0.02 \text{ (PDF)} \text{ fb}.
\end{equation*}
 
Combined with the measured signal strength, it results in an observed fiducial cross section of
\begin{equation*}
\sigma_{Z\gamma\text{EWK}}
= 0.77 ^{+0.34}_{-0.30} \text{ fb}
= 0.77 ^{+0.25}_{-0.23} \text{ (stat.)} ^{+0.22}_{-0.18} \text{ (syst.)} \text{ fb}.
\end{equation*}

\begin{table}[h!]
\begin{center}
\caption{Fiducial region definition.}
\label{table:FidVol}
\begin{tabular}{cc}
\toprule
Selections   &   Cut value  \\
\midrule
$\met$  &  $>120$~GeV \\
$E_\text{T}^{\gamma}$ & $>150$~GeV \\
Number of isolated photons & $N_{\gamma}$ = 1 \\
Photon isolation & $E_\text{T}^{\text{cone40}} < 0.022\pt + 2.45$~GeV, $\pt^{\text{cone20}}/\pt < 0.05$ \\
Number of jets & $N_\text{jets} \geq$ 2 with $p_{\mathrm{T}} >$ 50 GeV \\
Overlap removal & $\Delta R(\gamma,\mathrm{jet}) >$ 0.3 \\
Lepton veto & $N_{e}$ = 0, $N_{\mu}$ = 0 \\
$|\Delta\phi(\gamma, \vec{p}_{\mathrm{T}}^{\mathrm{\;miss}})|$ & $>0.4$ \\
$|\Delta\phi(j_\text{1}, \vec{p}_{\mathrm{T}}^{\mathrm{\;miss}})|$ & $>0.3$ \\
$|\Delta\phi(j_\text{2}, \vec{p}_{\mathrm{T}}^{\mathrm{\;miss}})|$ & $>0.3$ \\
$m_{jj}$ &  $> 300$ GeV \\
$\gamma$-centrality & $ < 0.6$ \\
\bottomrule
\end{tabular}
\end{center}
\end{table}
 
\begin{figure}
\centering
\subfloat[]{
\includegraphics[width=0.4\linewidth]{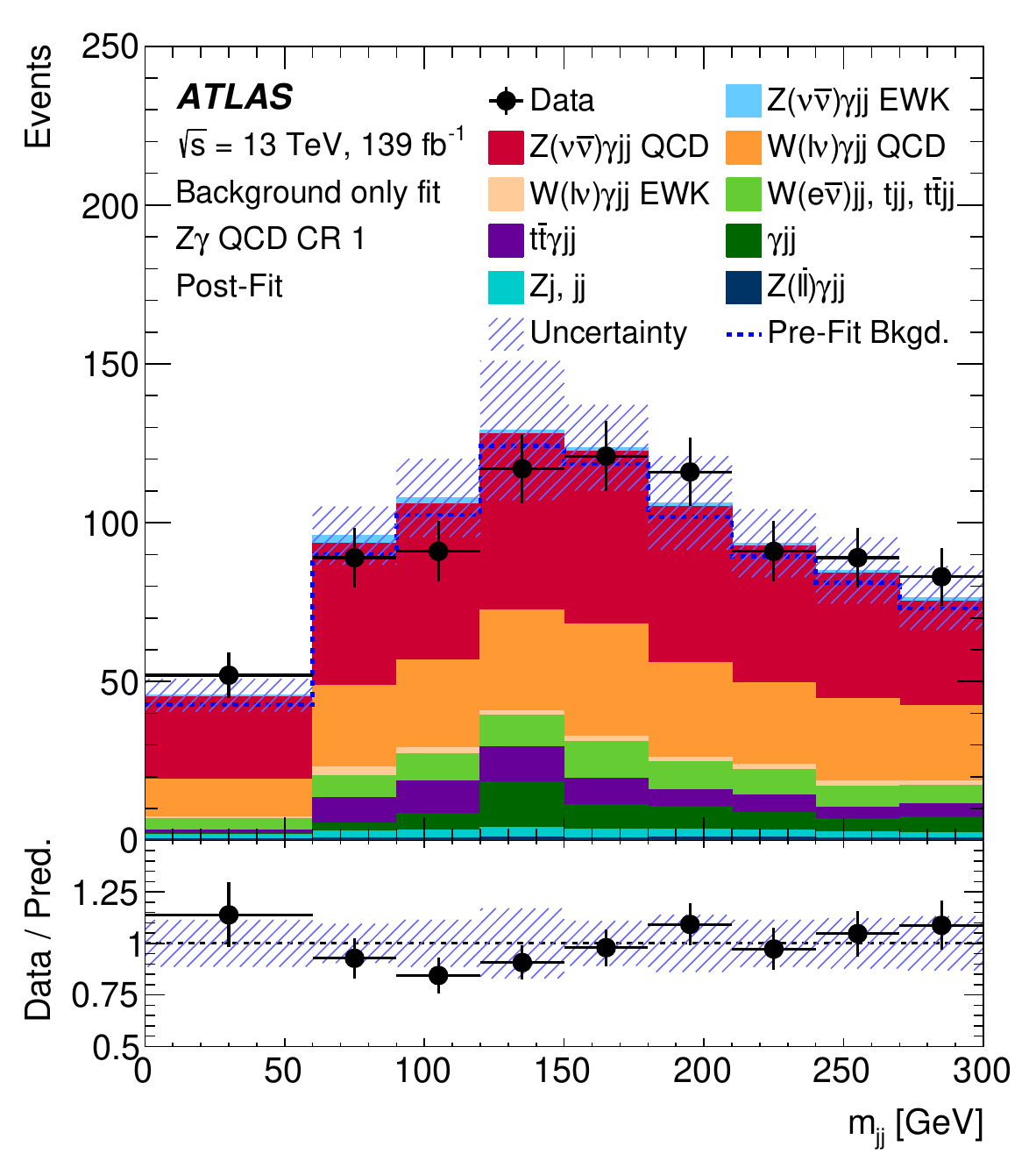}
\label{fig:observedPostFitZgQCD_1}
}
\subfloat[]{
\includegraphics[width=0.4\linewidth]{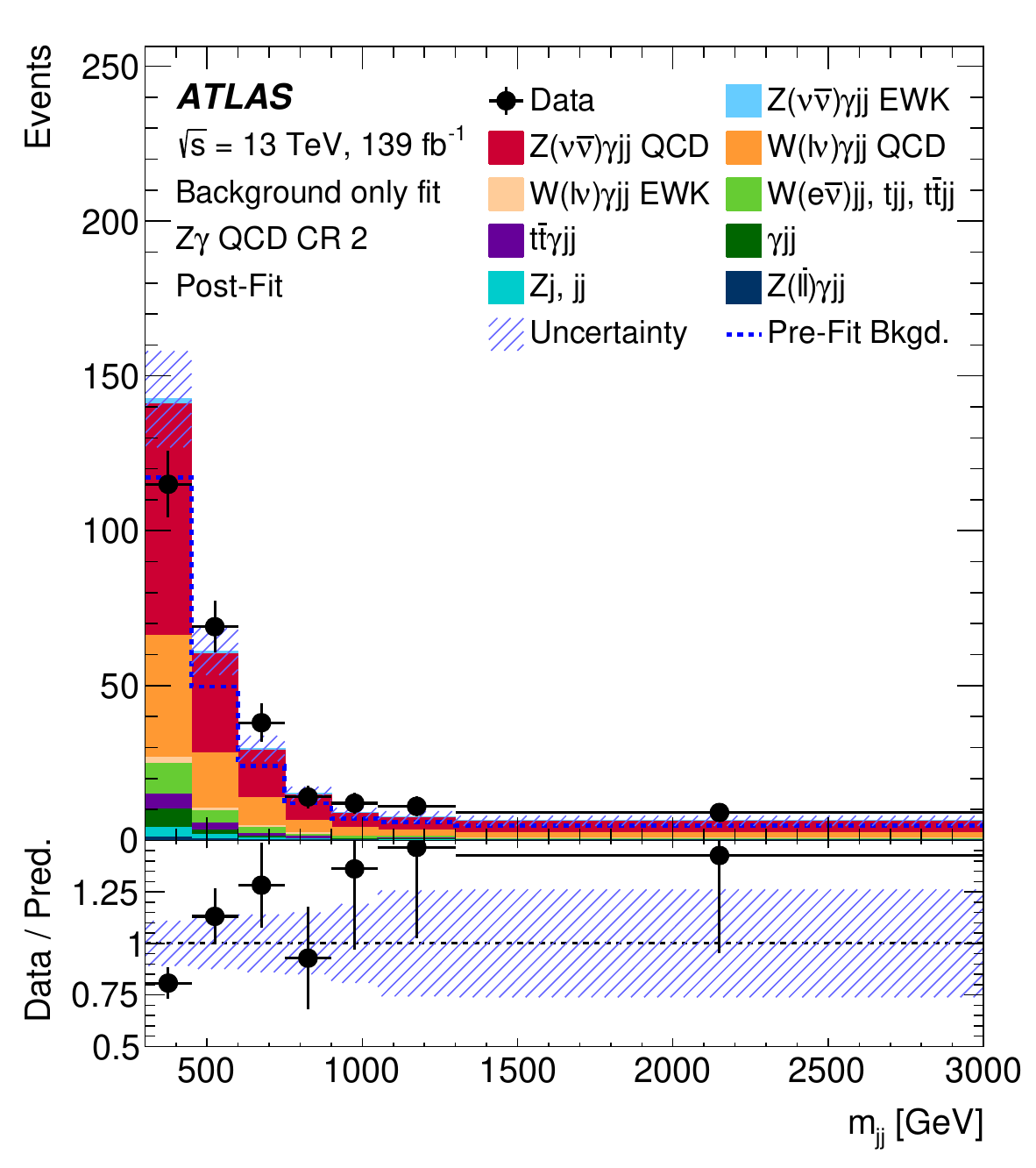}
\label{fig:observedPostFitZgQCD_2}
}
 
\subfloat[]{
\includegraphics[width=0.4\linewidth]{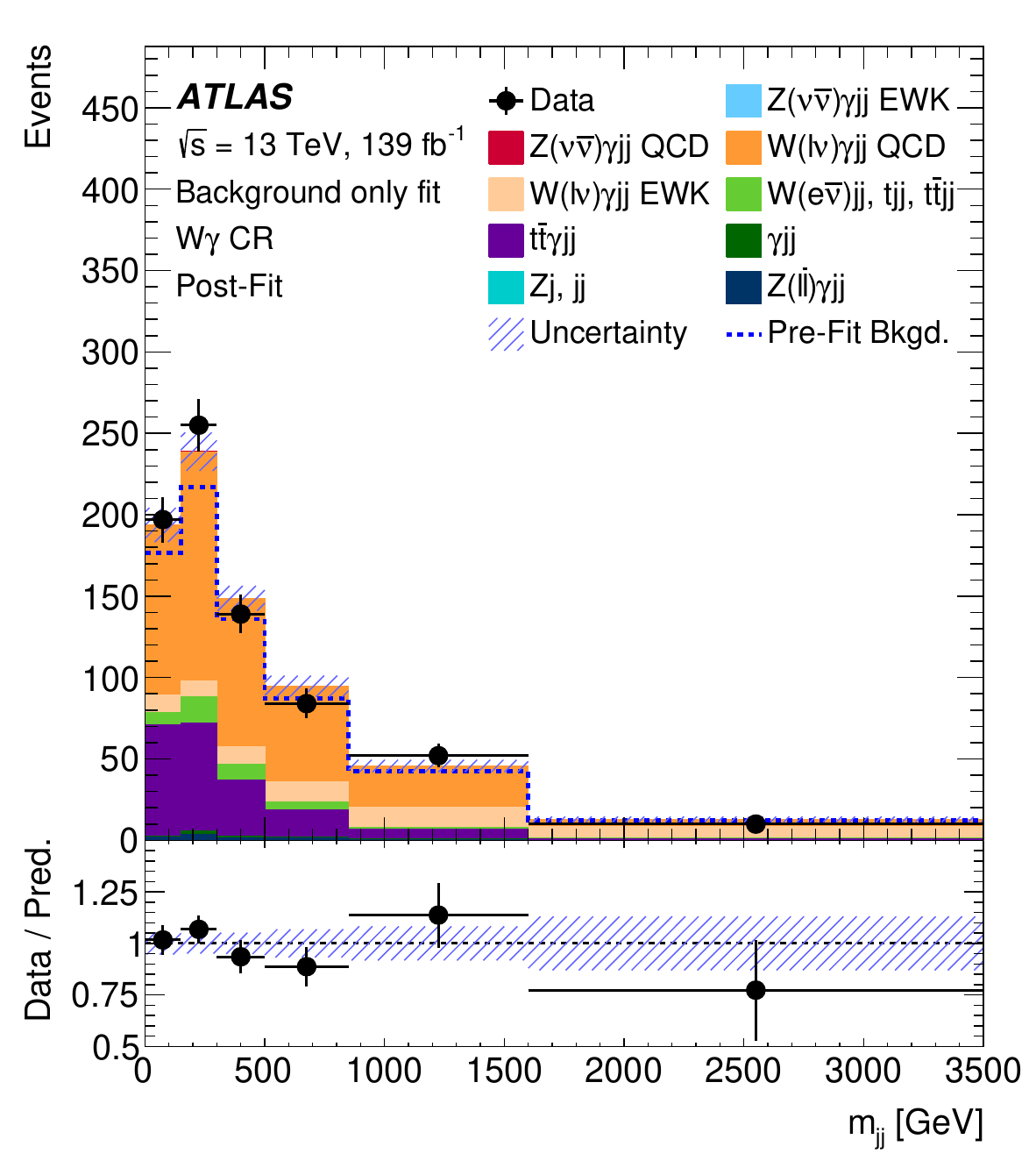}
\label{fig:observedPostFitWg}
}
\subfloat[]{
\includegraphics[width=0.4\linewidth]{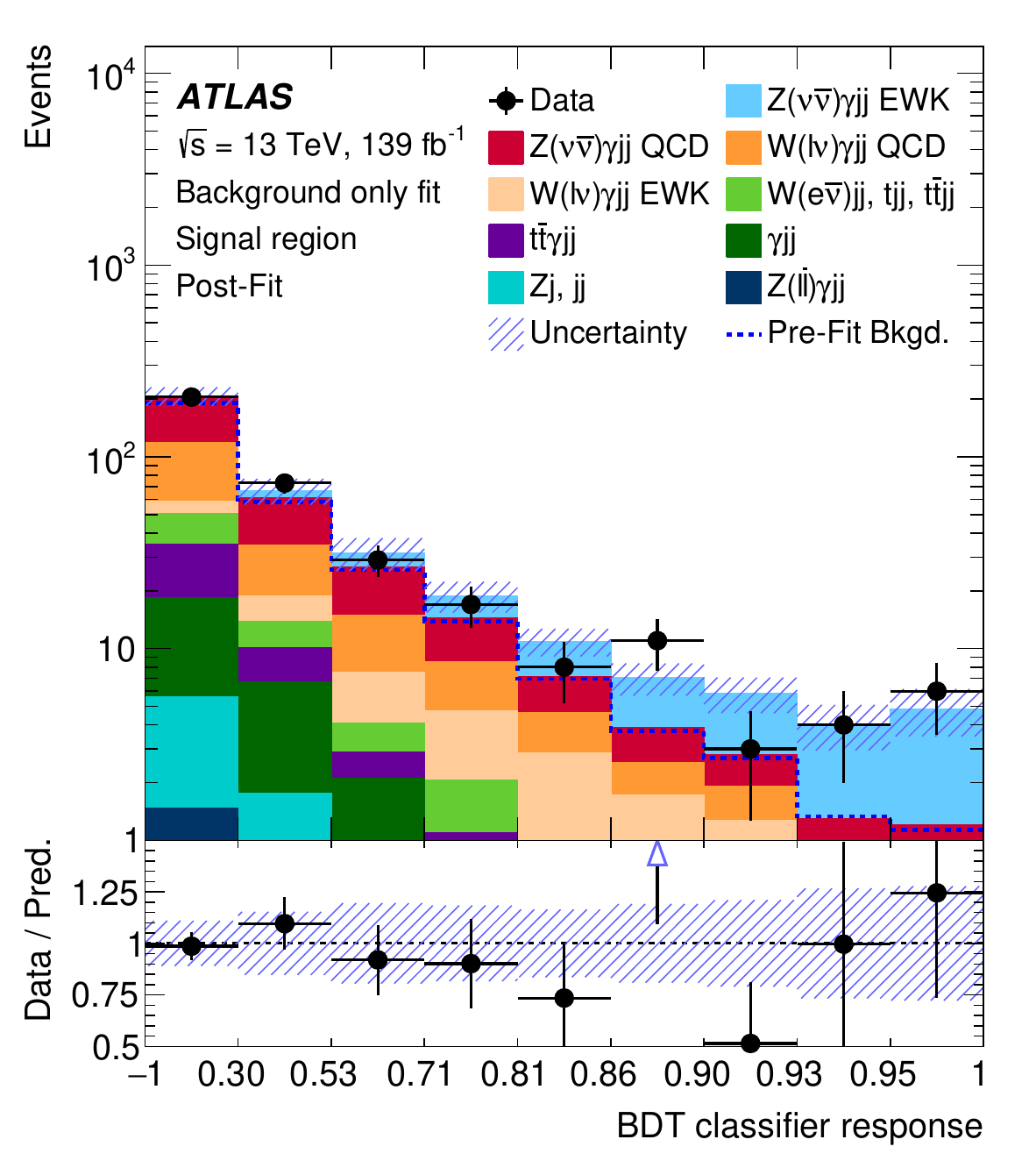}
\label{fig:observedPostFitSR}
}
\caption{The \mjj{} distributions for the \protect\subref{fig:observedPostFitZgQCD_1} $Z\gamma$ QCD CR\,1, \protect\subref{fig:observedPostFitZgQCD_2} $Z\gamma$ QCD CR\,2, and \protect\subref{fig:observedPostFitWg} $W\gamma$ CR, and the BDT classifier response distribution for the \protect\subref{fig:observedPostFitSR} SR after the fit in all regions. The BDT classifier response was remapped into equal-width bins for better representation. The dashed line shows the total background distribution before the fit. The vertical error bars on the data points correspond to the data's statistical uncertainty. Overflows are included in the last bin. The lower panel shows the ratio of observed to expected event yields. The uncertainty band corresponds to the combination of the statistical and systematic uncertainties obtained in the fit.}
\label{fig:observedPostFit}
\end{figure}

\begin{figure}
\centering
\includegraphics[width=0.6\linewidth]{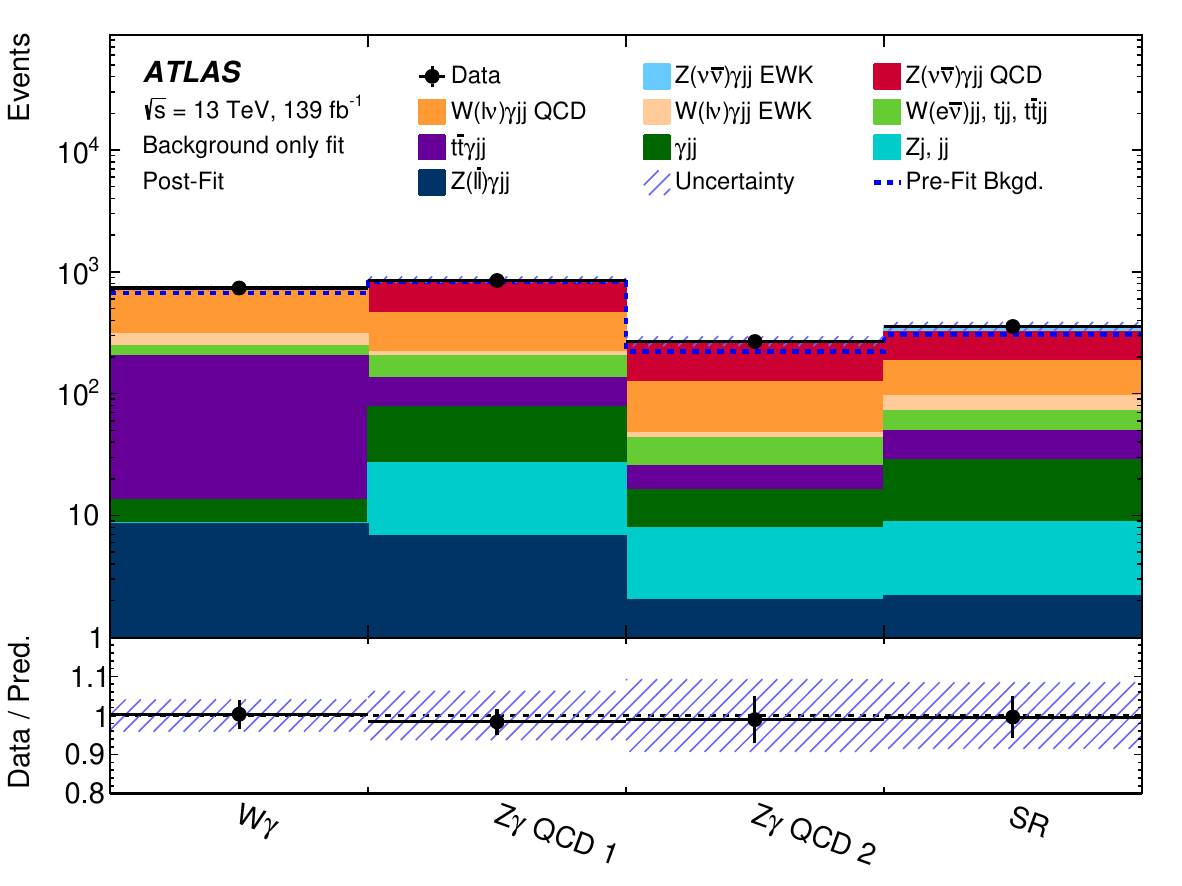}
 
\caption{Summary of the yield for processes in all regions, after the fit over all regions. The dashed line shows the total background distribution before the fit. The vertical error bars on the data points correspond to the data's statistical uncertainty. The lower panel shows the ratio of observed to expected event yields. The uncertainty band corresponds to the combination of the statistical and systematic uncertainties obtained in the fit.}
\label{fig:observedSummary}
\end{figure}

\Tab{\ref{table:eventYieldsPostFitUnblinded}} shows the observed and expected event yields of the signal and backgrounds in the SR and CRs after the fit is performed.
The post-fit \mjj{} and BDT classifier response distributions are shown in \Fig{\ref{fig:observedPostFit}}, and the summary plot for all of the regions is shown in \Fig{\ref{fig:observedSummary}}.
 
\begin{table}[h]
\centering
\caption{Observed  and expected event yields for the signal and all of the background processes considered in this analysis after the fit to the data in all of the regions. The uncertainty in the expected yield is the combination of statistical and systematic uncertainties obtained in the fit. The individual uncertainties can be correlated and do not necessarily add in quadrature to equal the total expected uncertainty.}
\label{table:eventYieldsPostFitUnblinded}
\begin{tabular}{l
r@{$\,\pm\,$}l
r@{$\,\pm\,$}l
r@{$\,\pm\,$}l
r@{$\,\pm\,$}l}
\toprule
& \multicolumn{2}{c}{$W\gamma$ CR} & \multicolumn{2}{c}{$Z\gamma$ QCD CR\,1} & \multicolumn{2}{c}{$Z\gamma$ QCD CR\,2} & \multicolumn{2}{c}{Signal region}\\
\midrule
$Z(\nu\bar{\nu})\gamma jj$~EWK   & 0.108 & 0.028 & 11.0 & 4.3 & 4.0 & 2.2 & 37 & 14 \\
$Z(\nu\bar{\nu})\gamma jj$~QCD   & 1.04 & 0.46 & 394 & 84 & 143 & 32 & 133 & 39 \\
$W(\ell\nu)\gamma jj$~QCD   & 425 & 63 & 237 & 71 & 76 & 24 & 91 & 30 \\
$W(\ell\nu)\gamma jj$~EWK   & 63 & 12 & 14.3 & 2.7 & 4.5 & 1.2 & 24.6 & 4.9 \\
$W(e\nu)jj$, $tjj$, $t\bar{t}jj$   & 39.8 & 2.5 & 70.1 & 4.1 & 17.9 & 1.3 & 22.5 & 1.5 \\
$t\bar{t}\gamma jj$   & 193 & 57 & 57 & 20 & 9.1 & 3.4 & 21.3 & 7.6 \\
$\gamma jj$   & 4.8 & 7.4 & 52 & 36 & 8 & 11 & 20 & 17 \\
$Zj$, $jj$   & 0.06 & 0.66 & 20 & 14 & 5.9 & 6.9 & 6.6 & 7.8 \\
$Z(\ell\bar{\ell})\gamma jj$   & 8.6 & 2.5 & 6.8 & 2.0 & 2.04 & 0.95 & 2.2 & 1.3 \\
\midrule
Total  & 735 & 30 & 863 & 54 & 271 & 25 & 357 & 30 \\
\midrule
Data   & \multicolumn{1}{l}{\phdo737} & & \multicolumn{1}{l}{\phd849} & & \multicolumn{1}{l}{\phd268} & & \multicolumn{1}{l}{\phd356} \\
\bottomrule
\end{tabular}
\end{table}
 
The breakdown of the impact of groups of systematic uncertainties on the cross-section measurement is shown in \Tab{\ref{table:systematicImpactBetter}}, with the theoretical uncertainties of the electroweak signal and the $Z(\nu\bar{\nu})\gamma jj$~QCD background having the largest impact.
 
\begin{table}
\centering
\caption{Impact of different components of the systematic uncertainty on the measured cross section, without taking into account the correlations. The impact is calculated by fixing the value of the corresponding nuisance parameters to the values obtained in the fit used to measure the cross section, performing the fit, estimating the signal strength uncertainty, subtracting its square from the square of the nominal uncertainty, and calculating the square root.}
\label{table:systematicImpactBetter}
\begin{tabular}{l
l@{\,/\,}l}
\toprule
Source of uncertainty & \multicolumn{2}{c}{$\Delta\sigma/\sigma~[\%]$}  \\
\midrule
\textbf{Experimental} & \multicolumn{2}{c}{} \\
\midrule
Jets & $-3.2$ & $+3.4$\\
Electrons and photons & $-0.3$ & $+1.7$\\
Muons & $-0.4$ & $+0.5$\\
\met{} & $-1.8$ & $+2.2$\\
Pile-up modelling & $-1.7$ & $+3.2$\\
Trigger efficiency & $-0.9$ & $+2.1$\\
Luminosity & $-1.2$ & $+2.6$\\
\midrule
\textbf{Theory} & \multicolumn{2}{c}{} \\
\midrule
$Z(\nu\bar{\nu})\gamma jj$~EWK/QCD interference &  $-0.6$ & $+2.6$\\
$Z(\nu\bar{\nu})\gamma jj$~EWK process & $-6$ & $+12$\\
$Z(\nu\bar{\nu})\gamma jj$~QCD process & $-15$ & $+16$\\
Other processes & $-5.3$ & $+7.7$\\
\midrule
\textbf{Other sources} &  \multicolumn{2}{c}{}\\
\midrule
Data-driven backgrounds & $-0.9$ & $+1.2$\\
Pile-up background & $-1.2$ & $+2.6$\\
$Z(\nu\bar{\nu})\gamma jj$~QCD \mjj{} modelling & $-4.4$ & $+4.4$\\
\bottomrule
\end{tabular}
\end{table}


\section{Combination with previous ATLAS measurement}
\label{sec:comb}

To increase the sensitivity, the measurement of electroweak \ensuremath{Z(\nu \bar{\nu}) \gamma jj}\xspace production presented in this paper is combined with the measurement from the previously published ATLAS observation of this process~\cite{EXOT-2021-17}. The analyses are statistically independent because their phase-space regions are orthogonal in $E_\mathrm{T}^{\gamma}$. This analysis requires $\et^{\gamma} > 150$~\GeV, while the previous ATLAS analysis requires $15 < \et^{\gamma} < 110$~\GeV.
 
The combined \ensuremath{Z(\nu \bar{\nu}) \gamma jj}\xspace EWK signal strength and its significance are extracted from a simultaneous profile likelihood fit. The fit includes all signal and control regions of both analyses and all corresponding systematic uncertainties. Various correlation schemes of jet energy and theoretical QCD scale uncertainties were tested and found to have negligible effect on the combined result. The observed~(expected) significance of the combined result with $\mu_{Z\gamma\text{EWK}}$ as the parameter of interest (POI) is $6.3 \sigma$~($6.6\sigma$). The fitted values of the signal strength ($\mu_{Z\gamma\text{EWK}}$) and background normalisations ($\mu_{Z\gamma\text{QCD}}$ and $\mu_{W\gamma}$) are listed in Table~\ref{tab:combination_pois}.
 
\begin{table} \centering
\caption{Fitted POI values for this analysis, the previous ATLAS analysis, and their combination. The second and third columns present the values obtained in the individual analyses. The fourth column presents the values obtained in the combination.  }
\begin{tabular}{l l@{ $\pm$ }l l@{ $\pm$ }l l@{ $\pm$ }l }
\toprule
& \multicolumn{6}{c}{Value} \\
\cmidrule(lr){2-7}
POI & \multicolumn{2}{l}{Current analysis}& \multicolumn{2}{l}{Ref.~\cite{EXOT-2021-17}}& \multicolumn{2}{l}{Combination}  \\
\midrule
$\mu_{Z\gamma\text{EWK}}$ & 0.78 & 0.33 & 1.03 & 0.25 & 0.96 & 0.19 \\
$\mu_{Z\gamma\text{QCD}}$ & 1.21 & 0.37 & 1.02 & 0.41 & 1.17 & 0.27 \\
$\mu_{W\gamma}$ & 1.02 & 0.22 & 1.01 & 0.20 & 1.01 & 0.13  \\
\bottomrule
\end{tabular}
 
\label{tab:combination_pois}
\end{table}
 
The difference in $\mu_{Z\gamma\text{EWK}}$ for the two analyses in combination is due to the lower data statistics in the signal region of the current analysis. The difference in $\mu_{Z\gamma\text{QCD}}$ for two analyses in combination is due to the different renormalisation and refactorisation scale correlation schemes used in the individual analyses. The most conservative correlation scheme that results in the highest expected uncertainty of the POI is used for both analyses to obtain the combination result.
 
The observed cross section using the combined signal strength is $9.2 \pm 2.0$~fb.
The fiducial region definition for this cross section is based on the one described in \Tab{\ref{table:FidVol}} with the following changes: the photon isolation and $\gamma$-centrality requirements were removed and the $\ET^{\gamma}$ threshold was lowered to 15~\GeV. This new, larger fiducial region includes the fiducial regions of both analyses used in the combination.
The predicted cross section extrapolated to this fiducial region using \VBFNLO is $9.6 \pm 1.0$~fb.


\section{Limits on anomalous quartic gauge couplings}
\label{sec:aqgc}

The results presented in Section~\ref{sec:sig_extraction} are used to set limits on anomalous QGCs via the VBS component of the measured electroweak process. 
New physics beyond the SM could induce anomalous QGCs, enhancing the \ensuremath{Z(\nu \bar{\nu}) \gamma jj}\xspace electroweak production cross section and modifying the kinematic distributions of the final-state bosons.
 
The effect of new physics introduced by aQGCs can be realised using an EFT~\cite{bib:eft} linearly parameterised by an effective Lagrangian as:
\begin{equation*}
\mathcal{L} = \mathcal{L}^\text{SM} + \displaystyle\sum\limits_i \frac{c_{i}}{\Lambda^{2}} \mathcal{O}_{i} + \displaystyle\sum\limits_j \frac{f_{j}}{\Lambda^{4}} \mathcal{O}_{j},
\end{equation*}
where $\mathcal{O}_{i}$ and $\mathcal{O}_{j}$ are dimension-6 or dimension-8 operators induced by integrating out the new degrees of freedom, while $c_{i}$ and $f_{j}$ represent the numerical coefficients that are meant to be derivable from a more complete high-energy theory. The $\Lambda$ term is a mass-dimension parameter associated with the energy scale of the new degrees of freedom that have been integrated out.
The \ensuremath{Z(\nu \bar{\nu}) \gamma jj}\xspace VBS process is sensitive to anomalous quartic and triple gauge couplings. Since the latter are well constrained in diboson production~\cite{STDM-2017-18}, they are not explored in this paper.
Among these higher-order operators, the dimension-8 ones are the lowest-dimensional operators inducing only quartic gauge-boson couplings without triple gauge-boson vertices. The impact of higher-dimensional operators is expected to be suppressed by more powers of the cut-off scale, $\Lambda$.
However the linear terms in the EFT coefficient of the process amplitude originating from the operators of some dimensions higher than eight, can be suppressed by the same or smaller power of $\Lambda$, compared with the quadratic term originating from eight-dimensional operators. Therefore, it is assumed that such contributions are suppressed by the dimensionless coupling constant, $f_{j}$. Moreover, higher-dimensional terms are currently not available and are thus not taken into account.
 
Two categories of dimension-8 operators contribute to the couplings in the studied final state: $\mathcal{O}_{TX}$ ($X=\text{0--9}$), constructed from the field-strength tensor; and $\mathcal{O}_{MX}$ ($X=\text{0--7}$), constructed from both the Higgs $\text{SU(2)}_\text{L}$ doublet derivatives and the field strength.
Seven operators are considered in this study, and the corresponding (Wilson) coefficients are: $f_{M0}/\Lambda^4$, $f_{M1}/\Lambda^4$, $f_{M2}/\Lambda^4$, representing $f_{MX}$ couplings, and $f_{T0}/\Lambda^4$, $f_{T5}/\Lambda^4$, $f_{T8}/\Lambda^4$ and $f_{T9}/\Lambda^4$, representing all types of $f_{TX}$ couplings. The sensitivity of the \ensuremath{Z(\nu \bar{\nu}) \gamma jj}\xspace EWK final state to these operators is competitive with other electroweak production modes. The last two of these couplings are unique and can be probed only by the neutral quartic vertices.
 
A clipping technique is introduced to preserve unitarity at very high parton centre-of-mass energies. Advantage of this unitarity restoring technique among the other ones is simplicity of application and further theoretical interpretation of the limits. The anomalous signal contribution is set to zero for $m_{Z\gamma} >$ $E_\text{c}$ (using particle-level information), where $E_\text{c}$ is a cut-off scale that is a free parameter. The chosen $E_\text{c}$ value is based on the unitarity bounds for a given limit value calculated from partial-wave unitarity constraints~\cite{PhysRevD.101.113003}.
 
Simulated \ensuremath{Z(\nu \bar{\nu}) \gamma jj}\xspace EWK events with non-zero EFT dimension-8 operator coefficients were generated by \MGNLO using decomposition of the process amplitude.
For the case of only one non-zero Wilson coefficient at a time, the squared amplitude is the following:
 
\begin{equation*}
|A|^2 = {|A_\text{SM} + f_j A_j|}^2 = {|A_\text{SM}|}^2 + f_j 2 \text{Re} (A_\text{SM}A_j^*)+ {f_j}^2 {|A_j|}^2,
\end{equation*}
 
where ${|A_\text{SM} + f_j A_j|}^2$ stands for the total amplitude squared with non-zero EFT parameter $f_j$, $A_\text{SM}$ is the Standard Model amplitude, $f_j 2 \text{Re} (A_\text{SM}A_j^*)$ is the amplitude of the interference between the SM and the EFT operator (the linear term of the process amplitude) and ${f_j}^2 {|A_j|}^2$ is the pure EFT operator contribution (quadratic term of the process amplitude). Individual samples using only one term at a time (SM, linear or quadratic terms) were generated for each operator.
To obtain the events at a given value of the EFT coefficient, the respective sample is multiplied by the appropriate value ($f_j$ or $f_j^2$).

\begin{figure}[h!]
\begin{center}
\includegraphics[width=0.80\linewidth]{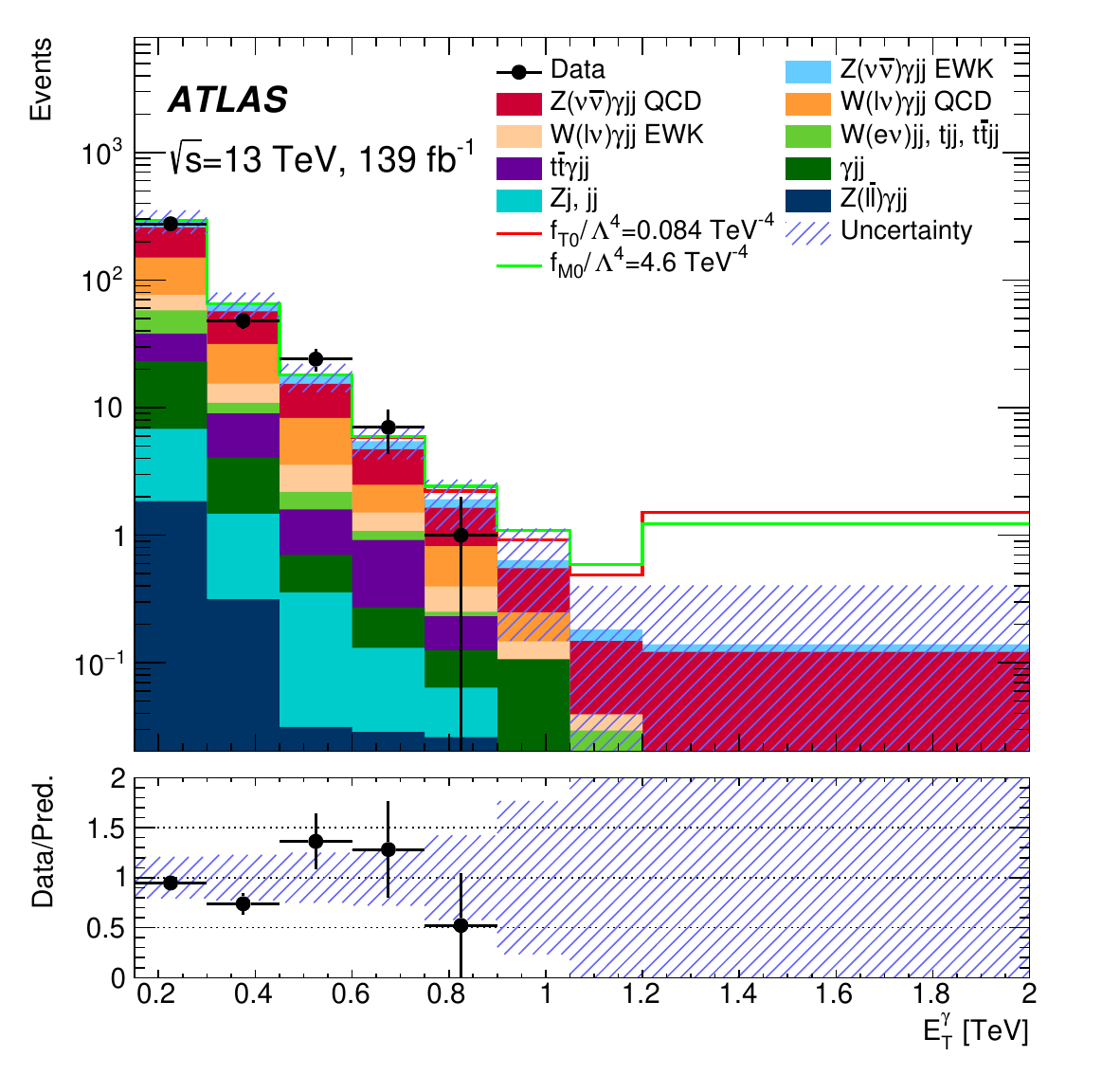}
\caption{The $\ET^{\gamma}$ distribution in the SR after the fit in the control regions. The red (green) line shows the expected number of events in the case of non-zero EFT coefficient $f_{T0}/\Lambda^4$ ($f_{M0}/\Lambda^4$) with the value shown in the legend. The vertical error bars on the data points correspond to the data statistical uncertainty. Overflows are included in the last bin. The lower panel shows the ratio of data to expected event yields. The uncertainty band corresponds to the combination of the MC statistical uncertainty and systematic uncertainties obtained in the fit.}
\label{fig:et_gamma_distr_aqgc}
\end{center}
\end{figure}

Limits on the dimension-8 operator coefficients are calculated using test statistics based on the profile likelihood ratio. The likelihood function is constructed as a product of a Poisson distribution and a Gaussian constraint term with nuisance parameters representing the sources of systematic uncertainty. Data event and predicted yields for the limit-setting procedure are taken from the signal region with additional optimisation of the $E_\mathrm{T}^{\gamma}$ threshold using expected confidence intervals for the EFT coefficients. The sensitivity to the aQGC is the strongest at high $E_\mathrm{T}^{\gamma}$, as it can be seen in Figure~\ref{fig:et_gamma_distr_aqgc}. Therefore, the constraints on the aQGC parameters come from a bin, constructed from the SR by the optimisation of additional $E_\mathrm{T}^{\gamma}$ threshold and corrected for background normalisations from the background-only fit.
Observed and expected 95\% CL intervals for the EFT coefficients are presented for two cases: when the clipping technique is not applied, and hence unitarity is not preserved, and when it is applied, and hence unitarity is preserved.
The results of the fit for the first and second cases are summarised in Tables~\ref{tab:observedExpected1DLimitsAQGC} and~\ref{tab:observedExpected1DLimitsAQGC_unit}, respectively.
The constraints are either competitive with or more stringent than those previously published by CMS~\cite{CMS-SMP-20-016,CMS-SMP-19-008,CMS-SMP-18-006}.

\begin{table}
\begin{center}
\caption{Observed and expected one-dimensional limits on dimension-8 aQGC coefficients. Limits are obtained by setting all aQGC coefficients except one to zero. Unitarity is not preserved.}
\label{tab:observedExpected1DLimitsAQGC}
\begin{tabular}{cll}
\toprule
Coefficient &  Observed limit [TeV$^{-4}$] & Expected limit [TeV$^{-4}$]  \\
\midrule
$f_{T0}/\Lambda^4$ & \quad$[-9.4, 8.4]\times10^{-2}$  & \quad$[-1.3, 1.2]\times10^{-1}$ \\
$f_{T5}/\Lambda^4$ & \quad$[-8.8, 9.9]\times10^{-2}$  & \quad$[-1.2, 1.3]\times10^{-1}$ \\
$f_{T8}/\Lambda^4$ & \quad$[-5.9, 5.9]\times10^{-2}$  & \quad$[-8.1, 8.0]\times10^{-2}$ \\
$f_{T9}/\Lambda^4$ & \quad$[-1.3, 1.3]\times10^{-1}$  & \quad$[-1.7, 1.7]\times10^{-1}$ \\
$f_{M0}/\Lambda^4$ & \quad$[-4.6, 4.6]$  & \quad$[-6.2, 6.2]$ \\
$f_{M1}/\Lambda^4$ & \quad$[-7.7, 7.7]$  & \quad$[-1.0, 1.0]\times10^1$ \\
$f_{M2}/\Lambda^4$ & \quad$[-1.9, 1.9]$  & \quad$[-2.6, 2.6]$ \\
\bottomrule
\end{tabular}
\end{center}
\end{table}
 
\begin{table}
\begin{center}
\caption{Observed and expected one-dimensional limits on dimension-8 aQGC coefficients in the region where unitarity is preserved. The cut-off scale $E_\text{c}$ in the simulation is given for each parameter. Limits are obtained by setting all aQGC coefficients except one to zero.}
\label{tab:observedExpected1DLimitsAQGC_unit}
\begin{tabular}{ccll}
\toprule
Coefficient & $E_\text{c}$ [\TeV]  & Observed limit [\TeV$^{-4}$] & Expected limit [TeV$^{-4}$]  \\
\midrule
$f_{T0}/\Lambda^4$ & 1.7  & \quad$[-8.7, 7.1]\times10^{-1}$ & \quad$[-8.9, 7.3]\times10^{-1}$ \\
$f_{T5}/\Lambda^4$ & 2.4  & \quad$[-3.4, 4.2]\times10^{-1}$ & \quad$[-3.5, 4.3]\times10^{-1}$ \\
$f_{T8}/\Lambda^4$ & 1.7  & \quad$[-5.2, 5.2]\times10^{-1}$ & \quad$[-5.3, 5.3]\times10^{-1}$ \\
$f_{T9}/\Lambda^4$ & 1.9  & \quad$[-7.9, 7.9]\times10^{-1}$ & \quad$[-8.1, 8.1]\times10^{-1}$ \\
$f_{M0}/\Lambda^4$ & 0.7  & \quad$[-1.6, 1.6]\times10^{2}$ & \quad$[-1.5, 1.5]\times10^{2}$ \\
$f_{M1}/\Lambda^4$ & 1.0  & \quad$[-1.6, 1.5]\times10^{2}$ & \quad$[-1.4, 1.4]\times10^{2}$ \\
$f_{M2}/\Lambda^4$ & 1.0  & \quad$[-3.3, 3.2]\times10^{1}$ & \quad$[-3.0, 3.0]\times10^{1}$ \\
\bottomrule
\end{tabular}
\end{center}
\end{table}
 
\begin{figure}[h!]
\begin{center}
\includegraphics[width=0.49\linewidth]{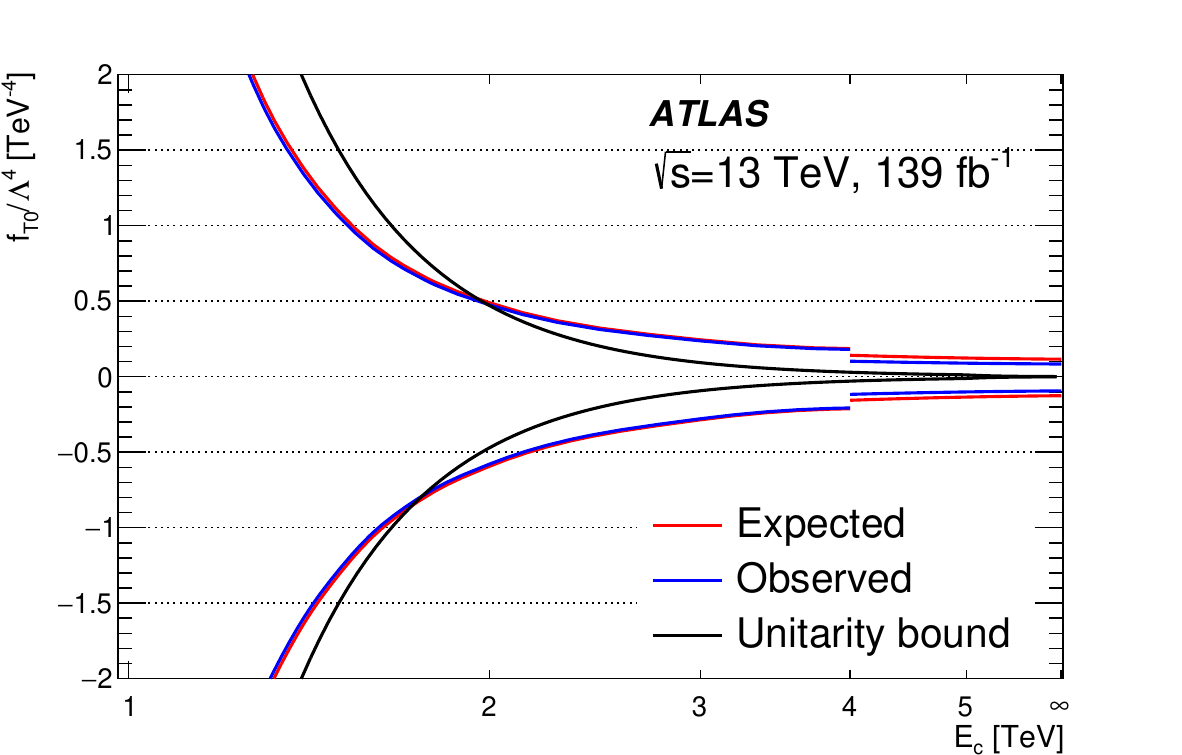}
\includegraphics[width=0.49\linewidth]{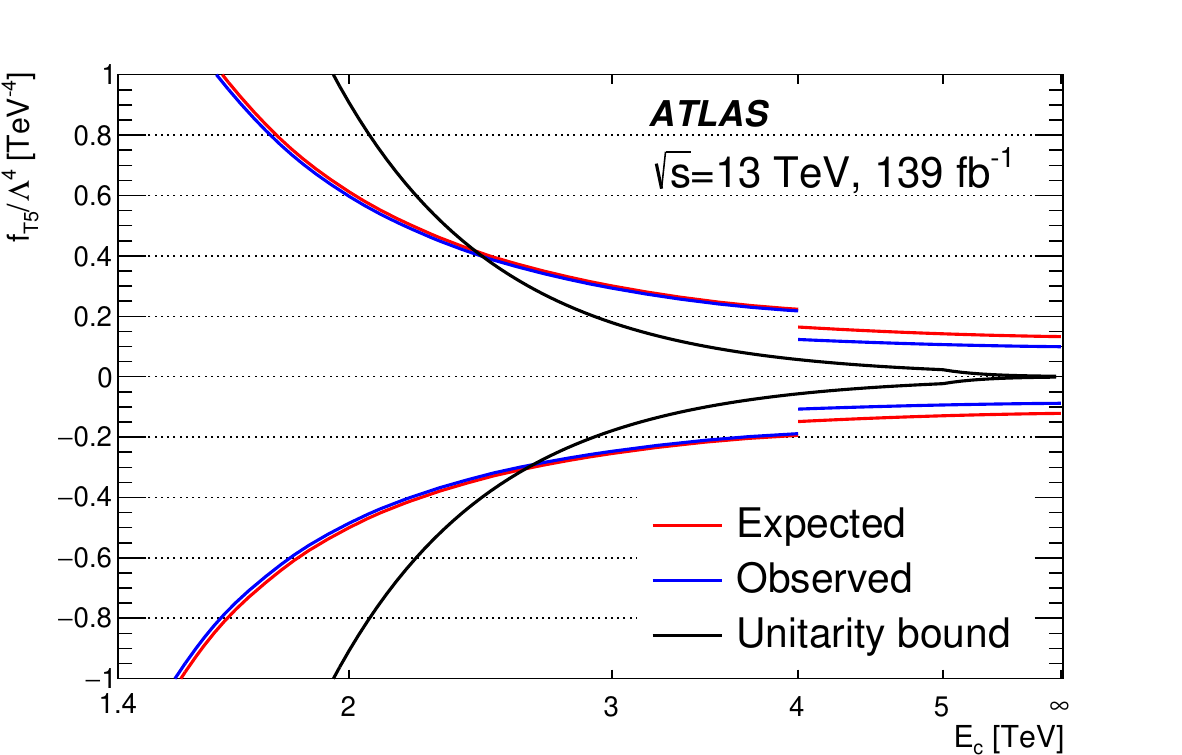}
\includegraphics[width=0.49\linewidth]{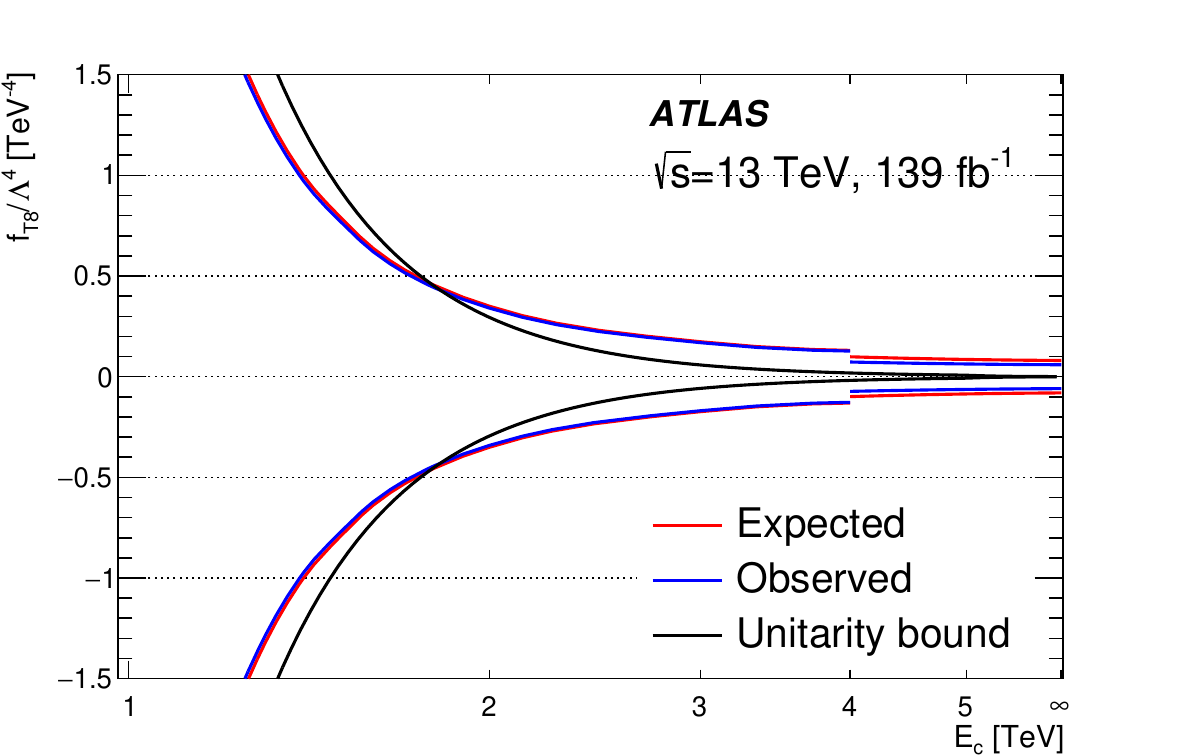}
\includegraphics[width=0.49\linewidth]{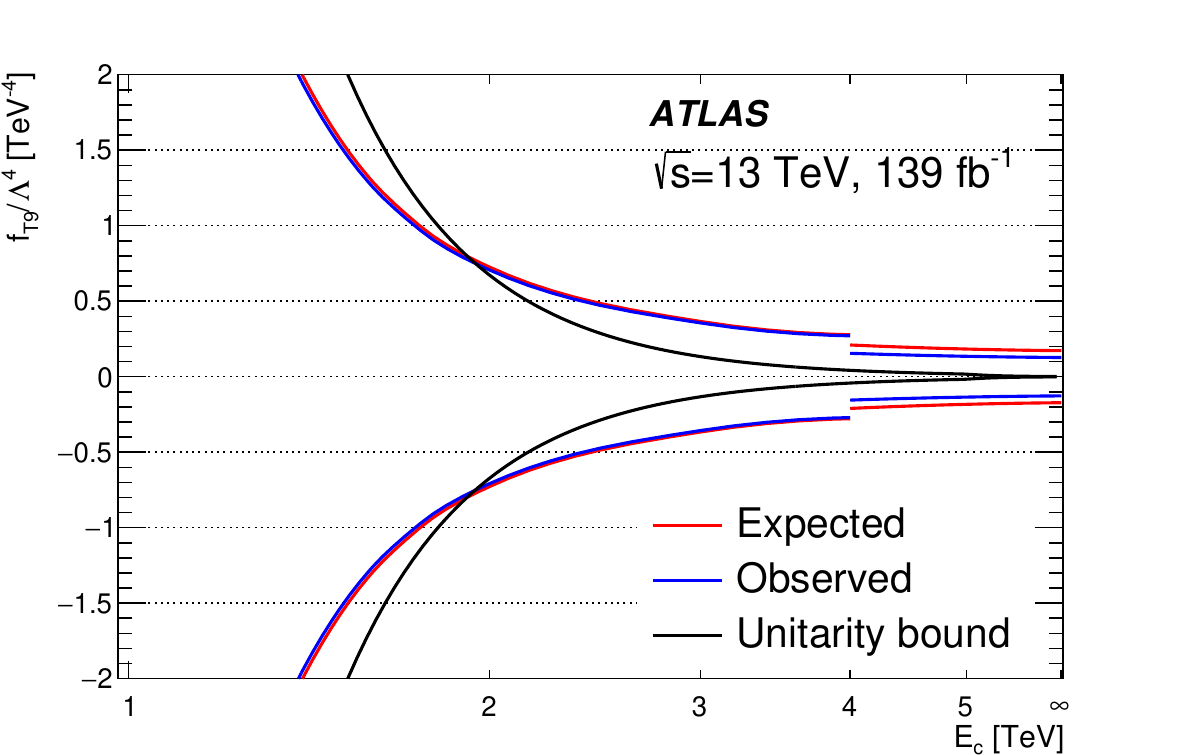}
\caption{Evolution of the expected (red line) and observed (blue line) limits versus $E_\text{c}$ values for $f_{T0}/\Lambda^4$, $f_{T5}/\Lambda^4$, $f_{T8}/\Lambda^4$ and $f_{T9}/\Lambda^4$. The unitarity bound is shown by the black line. The $E_\text{c}<4$~\TeV\ regime was obtained with $E_\mathrm{T}^{\gamma} > 600$~\GeV. The $E_\text{c}>4$~\TeV\ regime was obtained with $E_\mathrm{T}^{\gamma} > 900$~\GeV.}
\label{fig:limits_full_vs_Ec_1}
\end{center}
\end{figure}

\begin{figure}[h!]
\begin{center}
\includegraphics[width=0.49\linewidth]{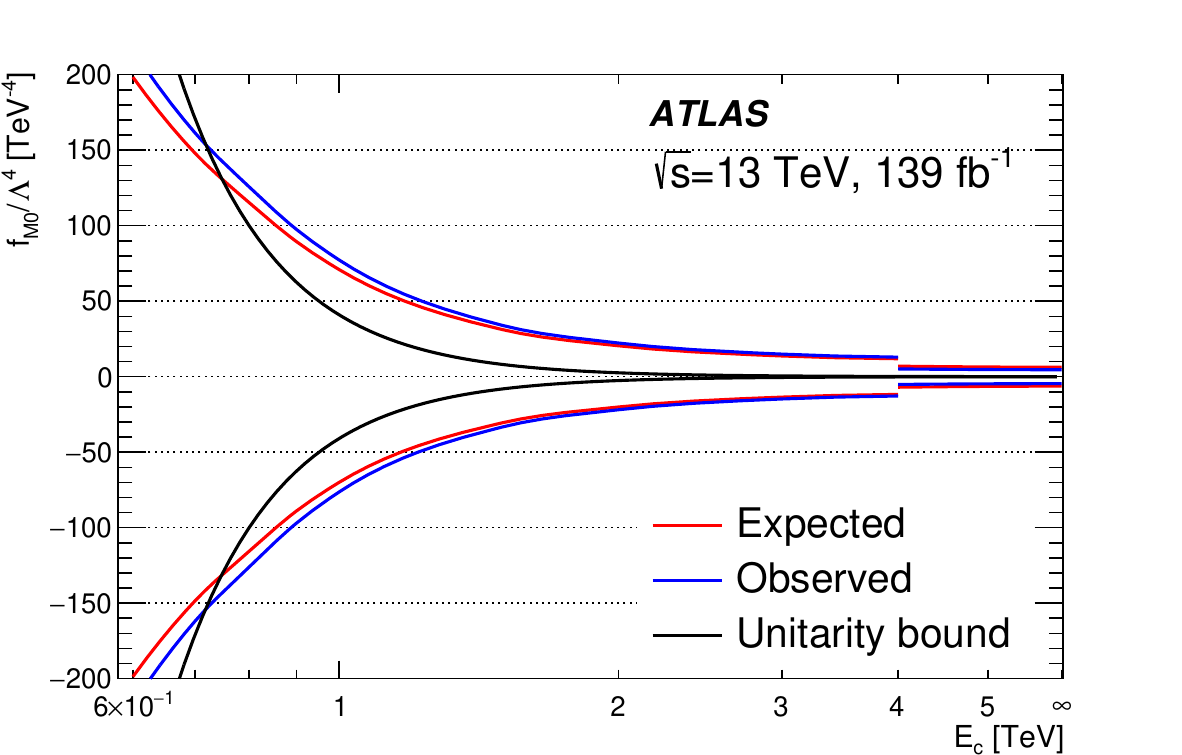}
\includegraphics[width=0.49\linewidth]{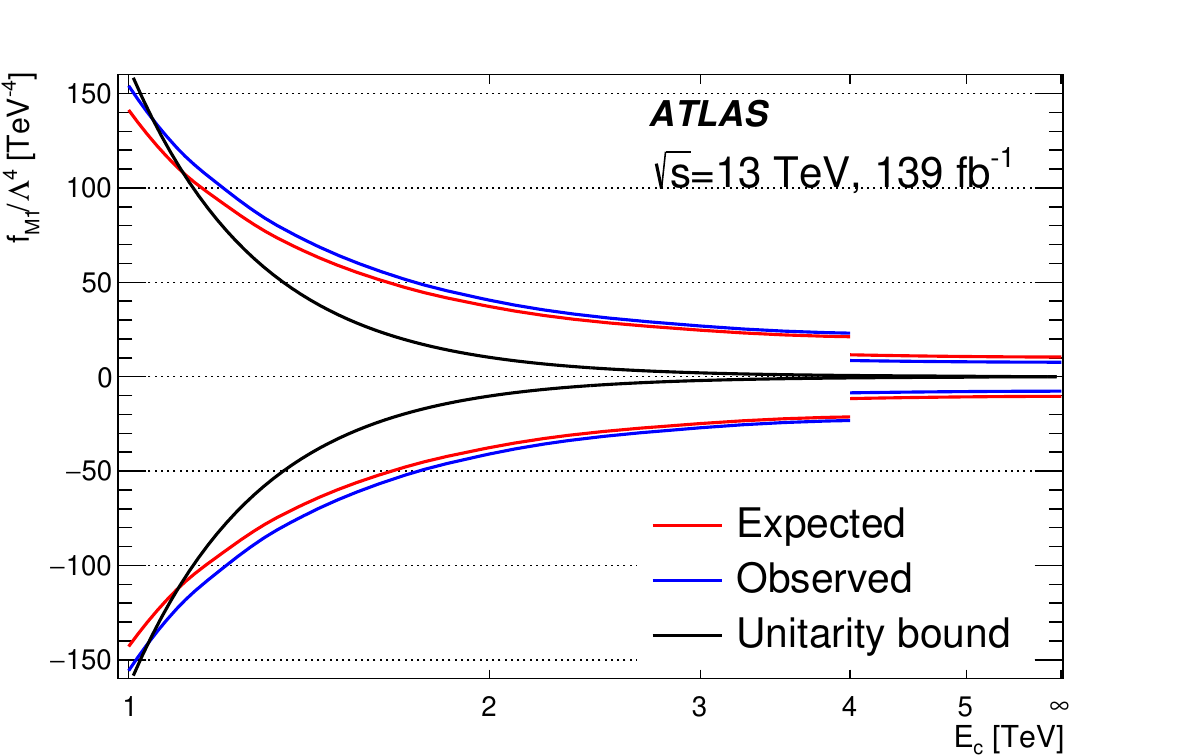}
\includegraphics[width=0.49\linewidth]{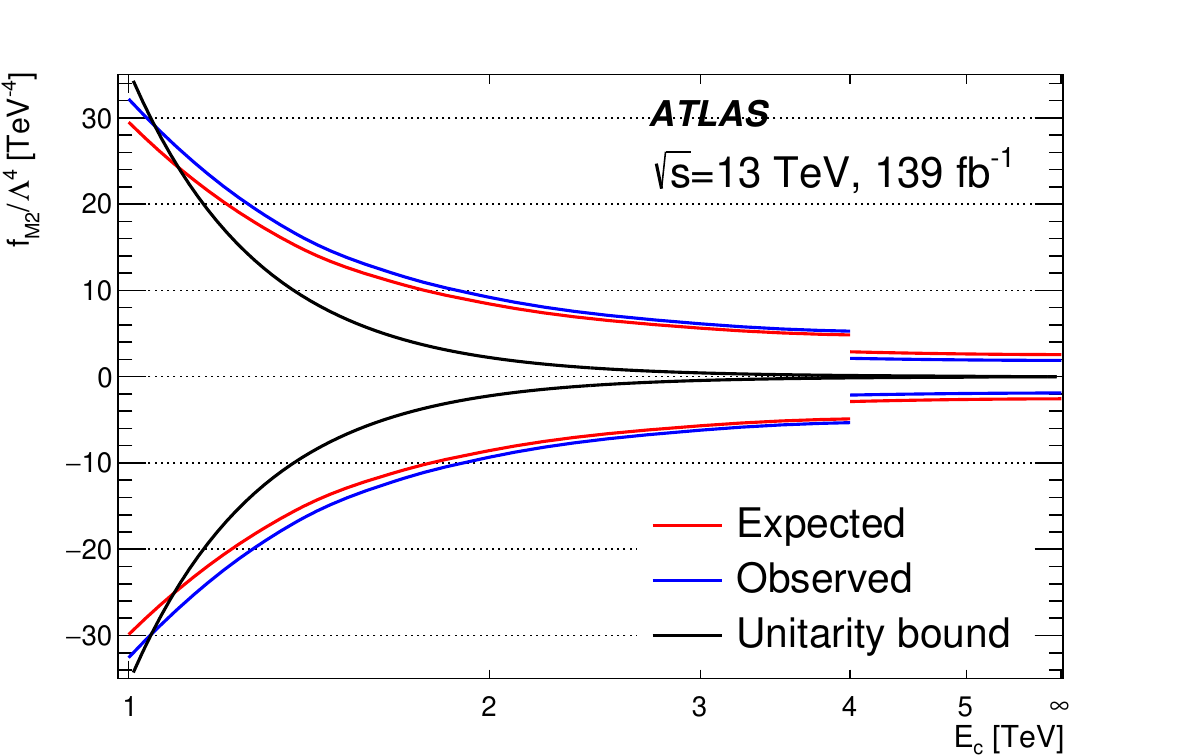}
\caption{Evolution of the expected (red line) and observed (blue line) limits versus $E_\text{c}$ values for $f_{M0}/\Lambda^4$, $f_{M1}/\Lambda^4$ and $f_{M2}/\Lambda^4$. The unitarity bound is shown by the black line. The $E_\text{c}<4$~\TeV\ regime was obtained with $E_\mathrm{T}^{\gamma} > 400$~\GeV. The $E_\text{c}>4$~\TeV\ regime was obtained with $E_\mathrm{T}^{\gamma} > 900$~\GeV.}
\label{fig:limits_full_vs_Ec_2}
\end{center}
\end{figure}

Illustrations of the limits' dependence on $E_\text{c}$ are given in Figures~\ref{fig:limits_full_vs_Ec_1} and~\ref{fig:limits_full_vs_Ec_2}.
The regime in which $E_\text{c}$ is less than 4~\TeV\ is obtained with an $E_\mathrm{T}^{\gamma}$ threshold of 600~\GeV\ (400~\GeV) for $f_{TX}$ ($f_{MX}$) couplings. The regime in which $E_\text{c}$ exceeds 4~\TeV\ is obtained with an $E_\mathrm{T}^{\gamma}$ threshold of 900~\GeV. The $E_\text{c}$ values for unitarised limits are obtained from the $E_\text{c}<4$~\TeV\ regime by taking the point before the one where the limits' dependence crosses the
unitarity bound. The infinity point in the figures indicates the limits obtained when the clipping technique is not applied, i.e.\ when $E_\text{c}=\infty$. The $E_\mathrm{T}^{\gamma}$ thresholds are chosen so that the analysis reaches its highest sensitivity. They are different for the unitarised and non-unitarised cases, since the cut-off removes events with high $E_\mathrm{T}^{\gamma}$.
 
\FloatBarrier
 
\section{Conclusion}
\label{sec:conclusion}

A measurement of the fiducial cross section for electroweak production of \ensuremath{Z(\nu \bar{\nu}) \gamma jj}\xspace in the region of $E_\mathrm{T}^{\gamma} > 150$~GeV is presented. Data from $\sqrt{s} = 13$~\TeV\ $pp$ collisions at the LHC were collected with the ATLAS detector during 2015--2018 and correspond to an integrated luminosity of $139$~fb$^{-1}$. The dominant backgrounds come from QCD mediated \ensuremath{Z(\nu \bar{\nu}) \gamma jj}\xspace and $W\gamma jj$ processes and these are evaluated using a simultaneous fit to data. Other significant backgrounds from \etogam and \jtogam misidentifications and \ETmiss mismeasurement are evaluated using data-driven techniques. The measurement uses the invisible decay mode of the gauge boson, $Z\to\nu\bar{\nu}$, and is performed in a fiducial phase space closely matching the detector acceptance.
 
The observed (expected) signal significance is 3.2$\sigma$~(3.7$\sigma$), which corresponds to evidence for this process in the given phase space used in the measurement. It was measured using a binned likelihood fit over the BDT classifier distribution.
The measured cross section is 0.77$^{+0.34}_{-0.30}$~fb, which is in agreement with SM predictions at NLO in perturbative QCD.
The cross sections and kinematics are quoted for the sum of the three neutrino flavours.
 
The results of this study are combined with those of the previously published ATLAS observation of this process to increase the sensitivity. This gives an observed (expected) signal significance of 6.3$\sigma$~(6.6$\sigma$).
 
Having found no significant deviations from SM predictions, the data are used to set limits on anomalous quartic gauge couplings. The limits are set on EFT dimension-8 operators $f_{T0}/\Lambda^4$, $f_{T5}/\Lambda^4$, $f_{T8}/\Lambda^4$, $f_{T9}/\Lambda^4$, $f_{M0}/\Lambda^4$, $f_{M1}/\Lambda^4$ and $f_{M2}/\Lambda^4$. These constraints are either competitive with or more stringent than those previously published by CMS. In particular, the constraints on the $f_{T5}/\Lambda^4$, $f_{T8}/\Lambda^4$ and $f_{T9}/\Lambda^4$ operators are significantly stronger than results previously published by ATLAS and CMS, based on either the full Run~2 dataset of $139$~fb$^{-1}$ or a partial dataset of $36$~fb$^{-1}$.


\section*{Acknowledgements}


We thank CERN for the very successful operation of the LHC, as well as the
support staff from our institutions without whom ATLAS could not be
operated efficiently.
 
We acknowledge the support of
ANPCyT, Argentina;
YerPhI, Armenia;
ARC, Australia;
BMWFW and FWF, Austria;
ANAS, Azerbaijan;
CNPq and FAPESP, Brazil;
NSERC, NRC and CFI, Canada;
CERN;
ANID, Chile;
CAS, MOST and NSFC, China;
Minciencias, Colombia;
MEYS CR, Czech Republic;
DNRF and DNSRC, Denmark;
IN2P3-CNRS and CEA-DRF/IRFU, France;
SRNSFG, Georgia;
BMBF, HGF and MPG, Germany;
GSRI, Greece;
RGC and Hong Kong SAR, China;
ISF and Benoziyo Center, Israel;
INFN, Italy;
MEXT and JSPS, Japan;
CNRST, Morocco;
NWO, Netherlands;
RCN, Norway;
MEiN, Poland;
FCT, Portugal;
MNE/IFA, Romania;
MESTD, Serbia;
MSSR, Slovakia;
ARRS and MIZ\v{S}, Slovenia;
DSI/NRF, South Africa;
MICINN, Spain;
SRC and Wallenberg Foundation, Sweden;
SERI, SNSF and Cantons of Bern and Geneva, Switzerland;
MOST, Taiwan;
TENMAK, T\"urkiye;
STFC, United Kingdom;
DOE and NSF, United States of America.
In addition, individual groups and members have received support from
BCKDF, CANARIE, Compute Canada and CRC, Canada;
PRIMUS 21/SCI/017 and UNCE SCI/013, Czech Republic;
COST, ERC, ERDF, Horizon 2020 and Marie Sk{\l}odowska-Curie Actions, European Union;
Investissements d'Avenir Labex, Investissements d'Avenir Idex and ANR, France;
DFG and AvH Foundation, Germany;
Herakleitos, Thales and Aristeia programmes co-financed by EU-ESF and the Greek NSRF, Greece;
BSF-NSF and MINERVA, Israel;
Norwegian Financial Mechanism 2014-2021, Norway;
NCN and NAWA, Poland;
La Caixa Banking Foundation, CERCA Programme Generalitat de Catalunya and PROMETEO and GenT Programmes Generalitat Valenciana, Spain;
G\"{o}ran Gustafssons Stiftelse, Sweden;
The Royal Society and Leverhulme Trust, United Kingdom.
 
The crucial computing support from all WLCG partners is acknowledged gratefully, in particular from CERN, the ATLAS Tier-1 facilities at TRIUMF (Canada), NDGF (Denmark, Norway, Sweden), CC-IN2P3 (France), KIT/GridKA (Germany), INFN-CNAF (Italy), NL-T1 (Netherlands), PIC (Spain), ASGC (Taiwan), RAL (UK) and BNL (USA), the Tier-2 facilities worldwide and large non-WLCG resource providers. Major contributors of computing resources are listed in Ref.~\cite{ATL-SOFT-PUB-2021-003}.


\printbibliography
 
\clearpage
 
\begin{flushleft}
\hypersetup{urlcolor=black}
{\Large The ATLAS Collaboration}

\bigskip

\AtlasOrcid[0000-0002-6665-4934]{G.~Aad}$^\textrm{\scriptsize 101}$,
\AtlasOrcid[0000-0002-5888-2734]{B.~Abbott}$^\textrm{\scriptsize 119}$,
\AtlasOrcid[0000-0002-7248-3203]{D.C.~Abbott}$^\textrm{\scriptsize 102}$,
\AtlasOrcid[0000-0002-1002-1652]{K.~Abeling}$^\textrm{\scriptsize 55}$,
\AtlasOrcid[0000-0002-8496-9294]{S.H.~Abidi}$^\textrm{\scriptsize 29}$,
\AtlasOrcid[0000-0002-9987-2292]{A.~Aboulhorma}$^\textrm{\scriptsize 35e}$,
\AtlasOrcid[0000-0001-5329-6640]{H.~Abramowicz}$^\textrm{\scriptsize 150}$,
\AtlasOrcid[0000-0002-1599-2896]{H.~Abreu}$^\textrm{\scriptsize 149}$,
\AtlasOrcid[0000-0003-0403-3697]{Y.~Abulaiti}$^\textrm{\scriptsize 116}$,
\AtlasOrcid[0000-0003-0762-7204]{A.C.~Abusleme~Hoffman}$^\textrm{\scriptsize 136a}$
% ...
\bigskip
\\

% ...
$^{29}$Physics Department, Brookhaven National Laboratory, Upton NY; United States of America.\\
% ...
$^{35}$$^{(a)}$Facult\'e des Sciences Ain Chock, R\'eseau Universitaire de Physique des Hautes Energies - Universit\'e Hassan II, Casablanca;$^{(b)}$Facult\'{e} des Sciences, Universit\'{e} Ibn-Tofail, K\'{e}nitra;$^{(c)}$Facult\'e des Sciences Semlalia, Universit\'e Cadi Ayyad, LPHEA-Marrakech;$^{(d)}$LPMR, Facult\'e des Sciences, Universit\'e Mohamed Premier, Oujda;$^{(e)}$Facult\'e des sciences, Universit\'e Mohammed V, Rabat;$^{(f)}$Institute of Applied Physics, Mohammed VI Polytechnic University, Ben Guerir; Morocco.\\
% ...
$^{55}$II. Physikalisches Institut, Georg-August-Universit\"{a}t G\"ottingen, G\"ottingen; Germany.\\
% ...
$^{101}$CPPM, Aix-Marseille Universit\'e, CNRS/IN2P3, Marseille; France.\\
$^{102}$Department of Physics, University of Massachusetts, Amherst MA; United States of America.\\
% ...
$^{116}$Department of Physics, New York University, New York NY; United States of America.\\
% ...
$^{119}$Homer L. Dodge Department of Physics and Astronomy, University of Oklahoma, Norman OK; United States of America.\\
% ...
$^{136}$$^{(a)}$Departamento de F\'isica, Pontificia Universidad Cat\'olica de Chile, Santiago;$^{(b)}$Millennium Institute for Subatomic physics at high energy frontier (SAPHIR), Santiago;$^{(c)}$Instituto de Investigaci\'on Multidisciplinario en Ciencia y Tecnolog\'ia, y Departamento de F\'isica, Universidad de La Serena;$^{(d)}$Universidad Andres Bello, Department of Physics, Santiago;$^{(e)}$Instituto de Alta Investigaci\'on, Universidad de Tarapac\'a, Arica;$^{(f)}$Departamento de F\'isica, Universidad T\'ecnica Federico Santa Mar\'ia, Valpara\'iso; Chile.\\
% ...
$^{149}$Department of Physics, Technion, Israel Institute of Technology, Haifa; Israel.\\
$^{150}$Raymond and Beverly Sackler School of Physics and Astronomy, Tel Aviv University, Tel Aviv; Israel.\\
% ...
$^{a}$ Also Affiliated with an institute covered by a cooperation agreement with CERN.\\
$^{b}$ Also at An-Najah National University, Nablus; Palestine.\\
$^{c}$ Also at Borough of Manhattan Community College, City University of New York, New York NY; United States of America.\\
$^{d}$ Also at Bruno Kessler Foundation, Trento; Italy.\\
$^{e}$ Also at Center for High Energy Physics, Peking University; China.\\
$^{f}$ Also at Centro Studi e Ricerche Enrico Fermi; Italy.\\
% ...

\end{flushleft}


\end{document}